%% file: main_draft.tex
\documentclass[a4paper,10pt,twocolumn]{article} %twocolumn, draft
\usepackage[utf8]{inputenc}
\usepackage{amsmath}
\usepackage{amssymb}

\usepackage[normalem]{ulem} % for strike-through with \sout{...}

\usepackage{graphicx}

\usepackage{natbib}
\bibpunct{(}{)}{;}{a}{}{,} % natbib cfg for j neurophysiol, might be also possible with makebst
\usepackage{xcolor}

\usepackage{siunitx} % for SI units, e.g., non-italic ``micro'' second
\usepackage{setspace} % for switching to doublespacing
\usepackage{geometry}
\usepackage{hyperref}
\hypersetup{pdfborder={0 0 0}} % links wont be visualized

\def\EE{\mathcal E} 
\def\II{\mathcal I}

\begin{document}

% \maketitle

\begin{flushleft}
{\Large
\textbf{How adaptation currents change threshold, gain and variability of neuronal spiking}}
% ... and synaptic inhibition ...
%Effects of adaptation currents and synaptic inhibition on threshold, gain and variability of neuronal spiking
%Control of gain and variability (of neuronal spiking) through adaptation currents
%Subthreshold and spike-triggered adaptation currents differentially affect gain and variability of neuronal spiking
\\
\vspace{0.25cm}
Josef Ladenbauer$^{1,2\ast}$, 
Moritz Augustin$^{1,2}$, 
Klaus Obermayer$^{1,2}$
\\
\vspace{0.25cm}
{\small
\textbf{1} Neural Information Processing Group, Technische Universit\"at Berlin, Berlin, Germany
\\
\textbf{2} Bernstein Center for Computational Neuroscience Berlin, Berlin, Germany 
\\
\vspace{0.25cm}
\textbf{$^\ast$Correspondence:} \\
Josef Ladenbauer\\
Technische Universit\"at Berlin\\
Department of Software Engineering and Theoretical Computer Science\\
Neural Information Processing\\
Marchstr. 23, MAR 5049\\
10587 Berlin, Germany\\
E-mail: jl@ni.tu-berlin.de}\\
\vspace{0.1cm}
\end{flushleft}

\input{section_abstract}

\input{section_introduction}

\input{section_methods}

\input{section_results}

\input{section_discussion}

\input{section_acknowledgements}

%\input{section_appendix}

%\bibliography{references}

\input{main_draft.bbl}
\input{section_figures}

\end{document}

%% file: section_abstract.tex
\begin{abstract}

%%% v1 %%%
%Neuronal inhibition is mediated by synaptic interaction or by intrinsic slow $\mathrm{K}^+$-currents %through different types of channels 
%which cause spike rate adaptation. 
%%% v2 %%%
%Many types of neurons exhibit spike rate adaptation, a form of intrinsic inhibition which is typically mediated by slow potassium currents.
%%through voltage-dependent, sodium-activated or calcium-activated channels. 
%% or:
Many types of neurons exhibit spike rate adaptation, mediated by intrinsic slow $\mathrm{K}^+$-currents, which effectively inhibit neuronal responses.
%
%These adaptation currents represent a form of inhibition which is cell-intrinsic in contrast to the one mediated by synaptic interaction. %These adaptation currents mediate intrinsic inhibition in contrast to the one caused by synaptic interaction.   
%Here we examine/ask how different types of adaptation currents ...
%How these adaptation currents, compared to synaptic inhibition,
%%or: compared to the one caused by synaptic interaction,   
%change important neuronal properties such as the relationship between synaptic input and spike rate output (I-O curve) as well as inter-spike interval (ISI) statistics, is not well understood. %These two neuronal properties have been found important for stimulus selectivity, coding and attention.
%
%%% v2 %%%
How these adaptation currents change the relationship between \emph{in-vivo} like fluctuating synaptic input, spike rate output and the spike train statistics, however, is not well understood.
%%% v1 %%%
%Whether these different types of inhibition change the relationship between synaptic input and spike rate output (I-O curve) as well as inter-spike interval (ISI) statistics in a similar way, and how, is not well understood.
%
%They contribute to frequency selectivity, coding and attention. These effects are likely caused by altering the relationship between synaptic input and spike rate output (I-O curve) as well as the inter-spike interval (ISI) statistics. %How the different types of adaptation current affect these properties however is not well understood.
%% 
%Here we provide a systematic study on (i) how the neuronal I-O curve and ISIs are altered by %subthreshold voltage-dependent as well as spike-dependent adaptation currents 
%different types of adaptation current for a wide range of input statistics %on the slow timescale 
%and (ii) how these changes compare to those induced by synaptic inhibition. % which can occur rapidly. %[Motivate inhibition].
%%
%%% v1 %%%
%Using analytical as well as numerical tools we systematically %describe how different types of adaptation currents and synaptic inhibition alter spike rates %the neuronal I-O curve 
%describe %or: compare and contrast 
%the effects of different types of adaptation currents and synaptic inhibition on spike rates %the neuronal I-O curve 
%and ISIs for a wide range of input statistics. 
%%% v2 %%%
In this computational study we %systematically investigate %or describe 
show that an adaptation current which primarily depends on the subthreshold membrane voltage changes the neuronal input-output relationship (I-O curve) subtractively, thereby increasing the response threshold, and decreases its slope (response gain) 
for low spike rates. %in presence of input fluctuations.
%When input fluctuations are large compared to the average input
%this adaptation current also reduces the slope of the I-O curve (the response gain). 
A spike-dependent adaptation current alters the I-O curve divisively, thus reducing the response gain. Both types of adaptation currents naturally increase the mean inter-spike interval (ISI), but they can affect ISI variability in opposite ways. A subthreshold current always causes an increase of variability while a spike-triggered current decreases high variability caused by fluctuation-dominated inputs and increases low variability when the average input is large. %i.e., the sensitivity of spiking variability to the mean input is reduced. %or: ...reduces high variability at low spike rates and increases low variability at high spike rates. 
The effects on I-O curves match those caused by synaptic inhibition in networks with asynchronous irregular activity, for which we find
subtractive and divisive changes caused by external and recurrent inhibition, respectively.  
%When comparing the effects on I-O curves to those caused by synaptic inhibition in a network showing asynchronous irregular activity, we find 
%subtractive and divisive changes caused by external and recurrent inhibition, respectively. 
%..., we find correspondences between subthreshold adaptation currents and external inhibition as well as spike-triggered adaptation currents and recurrent inhibition. 
Synaptic inhibition, however, always increases the ISI variability. %, regardless of its origin. 
%Analytically derived expressions demonstrate the robustness of these results.
% or: Analytical results demonstrate the robustness of these effects.
%
We analytically derive expressions for the I-O curve and ISI variability, which demonstrate the robustness of our results. Furthermore, we show how the biophysical parameters of slow $\mathrm{K}^+$-conductances %in biophysical models 
contribute to the two different types of adaptation currents and find that
%
%We further quantify how the conductance properties of different slow $\mathrm{K}^+$-currents %in biophysical models 
%contribute to the two types of adaptation and find that % ... and ... components of different biophysical %\textcolor{gray}{models of} 
%%slow $\mathrm{K}^+$-currents and find...
$\mathrm{Ca}^{2+}$-activated $\mathrm{K}^+$-currents are effectively captured by a simple spike-dependent description, while muscarine-sensitive or $\mathrm{Na}^+$-activated $\mathrm{K}^+$-currents show a dominant subthreshold component.

\end{abstract}

%% file: section_introduction.tex
\section*{Introduction}

Adaptation is a widespread phenomenon in nervous systems, providing flexibility to function under varying external conditions. At the single neuron level this can be observed as spike rate adaptation, a gradual decrease in spiking activity following a sudden increase in stimulus intensity. This type of intrinsic inhibition, in contrast to the one caused by synaptic interaction, is typically mediated by
slowly decaying somatic $\mathrm{K}^+$-currents which accumulate when the membrane voltage increases. A number of slow $\mathrm{K}^+$-currents with different activation characteristics have been identified. Muscarine-sensitive \citep{Brown1980,Adams1982} or $\mathrm{Na}^+$-dependent $\mathrm{K}^+$-channels activate at subthreshold voltage values \citep{Schwindt1989,Kim1998}, whereas $\mathrm{Ca}^{2+}$-dependent $\mathrm{K}^+$-channels activate at higher, suprathreshold values \citep{Brown1983,Madison1984,Schwindt1992}.  %\citep{Brown1983, Madison1984, Schwindt1992, Smith2002, Stocker2004}.
%through voltage-dependent low-threshold or calcium-activated high-threshold channels (Brown and Adams 1980; Madison and Nicoll 1984; Stocker 2004), which are susceptible to cholinergic modulation (McCormick 1992).  

%Depending on their activation properties Such adaptation currents can change the frequency selectivity %or: stimulus selectivity and coding strategy 
%of neurons in different ways and they are likely involved in attentional modulation of neuronal responses by acetylcholine \citep{Herrero2008, Soma2012, Soma2013, McCormick1992}.
Such adaptation currents, for example, mediate frequency selectivity of neurons \citep{Fuhrmann2002,Benda2005,Ellis2007}, where the preferred frequency depends on the current activation type \citep{Deemyad2012}. \textcolor{black}{They promote network synchronization \citep{Sanchez-Vives2000,Augustin2013,Ladenbauer2013a}} and are likely involved in the attentional modulation of neuronal response properties by acetylcholine \citep{Herrero2008,Soma2012,McCormick1992}.
% These effects could be explained by ...
It has been hypothesized that these complex effects are produced by changing the relationship between synaptic input and spike rate output (I-O curve) \citep{Deemyad2012,Benda2003,Soma2012,Reynolds2009}. 
For example, changing the I-O curve of a neuron subtractively sharpens stimulus selectivity, whereas a divisive change downscales the neuronal response but preserves selectivity (see \citep{Wilson2012} in the context of synaptic inhibition).
%
%For example, adaptation currents which primarily depend on the subthreshold membrane voltage ... mediate frequency selectivity of neurons \citep{Fuhrmann2002, Benda2005, Ellis2007}, where the preferred frequency depends on the current activation type (subthreshold or spike-triggered) \citep{Deemyad2012}. The two types of adaptation current might differentially affect coding \citep{Prescott2008} depending on how they change ISI statistics and frequency selectivity? and both types are likely involved in attentional modulation of neuronal responses by acetylcholine \citep{Herrero2008, Soma2012, Soma2013, McCormick1992}. It has been hypothesized that these effects are produced by changing the relationship between synaptic input and spike rate output (I-O curve) \citep{Deemyad2012, Soma2012, Reynolds2009} and inter-spike interval (ISI) statistics \citep{Prescott2008}. 
%Moreover, a spike-dependent adaptation currents might promote spike-rate coding for slow inputs by regularizing the spike train while for a subthreshold adaptation current this is less clear. 
It was also suggested, that adaptation currents affect the neural code via their effect on the inter-spike interval (ISI) statistics \citep{Prescott2008}. 
%depending on how these currents affect the ISI statistics they might change the neuronal coding strategy ().
%How different types of adaptation currents change the neuronal I-O curve and ISI statistics for fluctuating inputs, however, is not well understood. 
%
%Therefore, it is important to understand how the different types of adaptation currents change the neuronal I-O curve and ISI statistics.
So far, effects of adaptation currents on I-O curves have been studied considering constant current inputs disregarding input fluctuations \citep{Prescott2008,Deemyad2012} and it has remained unclear how different types of adaptation currents affect ISI variability. 
%On the other hand, these properties can also be modulated by synaptic inhibition. For example, some types of inhibitory neurons alter the I-O curve of their target subtractively, thereby sharpening stimulus selectivity, while others cause a divisive change which downscales the neuronal response but preserves selectivity \citep{Wilson2012}.  
%
%In this way, synaptic inhibition can also shape neuronal tuning properties and plays an important role in gain control \citep{Isaacson2011, Silver2010}.
%
%These effects could be explained by changes of the relationship between synaptic input and spike rate output (I-O curve) \citep{Deemyad2012, Soma2012, Reynolds2009, Isaacson2011} inter-spike interval (ISI) statistics \citep{Prescott2008}.  %However, a detailed study on how different types of adaptation currents in comparison to synaptic inhibition alter the neuronal I-O curve and ISIs, is still lacking.
%
Therefore, in this contribution we systematically examine how voltage-dependent subthreshold and spike-dependent adaptation currents change neuronal I-O curves as well as the ISI distribution for typical \emph{in-vivo} like input statistics, % how these changes compare to those induced by synaptic inhibition in a network setting and (iii) 
and how the biophysical parameters of slow $\mathrm{K}^+$-conductances contribute to the two types of adaptation currents.

We address these questions by studying spike rates and ISI distributions of model neurons with subthreshold and spike-triggered adaptation currents, subject to fluctuating \emph{in-vivo} like inputs, and we compare the results to those induced by synaptic inhibition.
Specifically, we use the adaptive exponential integrate-and-fire (aEIF) neuron model \citep{Brette2005}, which has been shown to perform well in predicting the subthreshold properties \citep{Badel2008} and spiking activity \citep{Jolivet2008,Pospischil2011} of cortical neurons. %In addition, we use its simplified version, the adaptive perfect integrate-and-fire (aPIF) model, for which we analytically derive explicit expressions describing the I-O curve and spiking variability. 
%To analytically demonstrate the changes of I-O curves and spiking variability we derive explicit expressions for these properties 
%%or: for the steady-state spike rate and ISI coefficient of variation
%using the adaptive perfect integrate-and-fire (aPIF) neuron model, a simplified version of the aEIF model.
To analytically demonstrate the changes of I-O curves and ISI variability we derive explicit expressions for these properties based on the simpler perfect integrate-and-fire neuron model \textcolor{black}{(see, e.g., \citep{Gerstein1964})} with adaptation (aPIF).
%We analytically derive explicit expressions describing the I-O curve and spiking variability for the adaptive perfect integrate-and-fire (aPIF) neuron model, a simplified version of the aEIF model.
%Both neuron models allow to separately study the effects of subthreshold and spike-triggered adaptation currents. 
%
Finally, using a detailed conductance-based neuron model we quantify the subthreshold and spike-triggered components of various slow $\mathrm{K}^+$-currents and compare the effects of specific $\mathrm{K}^+$-channels on the I-O curve and ISI variability.
%Furthermore, we examine how the conductance properties of different slow $\mathrm{K}^+$-currents in a detailed biophysical neuron model contribute to a simple subthreshold or spike-dependent type of adaptation current  and compare their effects on the I-O curve and ISI variability. %and we compare the effects of the biophysical ($\mathrm{K}^+$-) and the simple (adaptation) currents.
%contribute to the two types of adaptation
%quantify how the two types of a simple adaptation current relate to different types of slow $\mathrm{K}^+$-currents using a detailed conductance-based neuron model and we compare the effects of the biophysical ($\mathrm{K}^+$-) and the simple (adaptation) currents.

%% file: section_methods.tex
\section*{Materials and Methods} 
%- aEIF terms\\
%- synaptic inputs, mu and sigma\\
%- adiabatic approximation\\
%- FP equation, p(V,t), r(t) \\
%- ISI distribution\\
%- Note on 2 population network motif\\
%%- HH type (TMS) neuron with noisy conductance inputs (OU noise)\\ 
%- Fitting adaptation parameters in aEIF model to HH model\\
%
%First, we describe the IF model neuron which includes an adaptation current with subthreshold and spike-triggered components and a noisy synaptic current. 
%Specifically, we use the adaptive exponential integrate-and-fire (aEIF) neuron model \citep{Brette2005}, which has been shown to perform well in predicting the spike times \citep{Jolivet2008} and post-stimulus time histograms \citep{Pospischil2011} of cortical neurons. 
%Based on this model, we then derive the Fokker-Planck equation in the limit of a large adaptation timescale to obtain the membrane voltage distribution, spike rate and ISI distribution of a single neuron receiving synaptic inputs. 
%We simplify the aEIF model to an adaptive perfect integrate-and-fire (aPIF) model for which we analytically derive explicit expressions for the steady-state spike rate and ISI coefficient of variation.
%Next, we calculate these quantities for an aEIF neuron that belongs to a network and receives external as well as recurrent synaptic inputs. % we describe how these quantities are calculated for ... / we consider that ...
%Finally, we describe the detailed neuron model with different biophysical slow $\mathrm{K}^+$-currents using the Hodgkin-Huxley formalism and quantify their subthreshold and spike-triggered components. \\

\noindent \textbf{aEIF neuron with noisy input current}\\
%We consider an adaptive exponential integrate-and-fire (aEIF) model neuron \citep{Brette2005} receiving synaptic input currents. 
We consider an aEIF model neuron receiving synaptic input currents.
The subthreshold dynamics of the membrane voltage $V$ is given by % maybe block the model eqs. instead
\begin{equation}
C \frac{dV}{dt} = I_{\mathrm{ion}}(V) + I_{\mathrm{syn}}(t), 
\label{membrane_eq}
\end{equation}
where the capacitive current through the membrane with capacitance $C$ equals the sum of ionic currents $I_{\mathrm{ion}}$ and the synaptic current $I_{\mathrm{syn}}$. Three ionic currents are taken into account,
\begin{equation}
I_{\mathrm{ion}}(V) :=  -g_{\mathrm{L}} (V-E_{\mathrm{L}}) + g_{\mathrm{L}} \, \Delta_{\mathrm{T}} \, 
%e^{\tfrac{V-V_{\mathrm{T}}}{\Delta_{\mathrm{T}}}} - w.
\exp \left( \frac{V-V_{\mathrm{T}}}{\Delta_{\mathrm{T}}} \right) - w.
\label{ionic_currents_eq}
\end{equation}
The first term on the right-hand side describes the leak current with conductance $g_{\mathrm{L}}$ and reversal potential $E_{\mathrm{L}}$. The exponential term with threshold slope factor $\Delta_{\mathrm{T}}$ and effective threshold voltage $V_{\mathrm{T}}$ approximates the fast $\mathrm{Na}^+$-current at spike initiation, assuming instantaneous activation of $\mathrm{Na}^+$-channels \citep{Fourcaud-Trocme2003}. $w$ is the adaptation current which reflects a slow $\mathrm{K}^+$-current. It evolves according to
\begin{equation}
\tau_w \frac{dw}{dt} = a(V-E_w) -w,
\label{w_current_eq}
\end{equation}
% 
%on a timescale given by ...
with adaptation time constant $\tau_w$. Its strength depends on the subthreshold membrane voltage via conductance $a$. $E_w$ denotes its reversal potential.
When $V$ increases beyond $V_{\mathrm{T}}$, a spike is generated due to the exponential term in eq.~\eqref{ionic_currents_eq}. The downswing of the spike is not explicitly modelled, instead, when $V$ reaches a value $V_{\mathrm{s}} \geq V_{\mathrm{T}}$, the membrane voltage is reset to a lower value $V_{\mathrm{r}}$. At the same time, the adaptation current $w$ is incremented by a value of $b$, implementing the mechanism of spike-triggered adaptation. Immediately after the reset, $V$ and $w$ are clamped for a refractory period $T_{\mathrm{ref}}$, and subsequently governed again by eqs.~\eqref{membrane_eq}--\eqref{w_current_eq}.

The aEIF model can reproduce a wide range of neuronal subthreshold dynamics \citep{Touboul2008} and spike patterns \citep{Naud2008}. 
%It has been shown to perform well in predicting the spike times \citep{Jolivet2008} and post-stimulus time histograms \citep{Pospischil2011} of cortical neurons.  
%Importantly, the effects of subthreshold and spike-triggered adaptation on response properties in the aEIF model well match the effects of M and AHP $\mathrm{K}^+$-currents in a detailed biophysical neuron model, respectively [Ladenbauer2012]. % <-- we actually want to show this relationship here
We selected the following parameter values to model cortical neurons: 
${C=\SI{1}{\micro\farad}/\mathrm{cm}^2}$, 
${g_{\mathrm{L}}=\SI{0.05}{\milli\siemens}/\mathrm{cm}^2}$,
${E_{\mathrm{L}}=\SI{-65}{\milli\volt}}$,
${\Delta_{\mathrm{T}}=\SI{1.5}{\milli\volt}}$,
${V_{\mathrm{T}}=\SI{-50}{\milli\volt}}$,
${\tau_w = \SI{200}{\milli\second}}$,
${E_w=\SI{-80}{\milli\volt}}$, \linebreak
${V_{\mathrm{s}}=\SI{-40}{\milli\volt}}$,
${V_{\mathrm{r}}=\SI{-70}{\milli\volt}}$ and
${T_{\mathrm{ref}}=\SI{1.5}{\milli\second}}$
\citep{Badel2008,Destexhe2009,Wang2003}. 
The adaptation parameters $a$ and $b$ were varied within reasonable ranges, ${a \in  [0,\,0.06]}$~mS/$\mathrm{cm}^2$, ${b \in [0,\,0.3]}$~$\SI{}{\micro\ampere}/\mathrm{cm}^2$. % \textbf{see fits}. 
%For inhibitory neurons adaptation was neglected (${a=b=0}$) since it was found to be weak in fast spiking interneurons compared to pyramidal cells \citepp{LaCamera2006}. 
 
The synaptic input consists of a mean $\mu(t)$ and a fluctuating part given by a Gaussian white noise process $\eta(t)$ with $\delta$-autocorrelation and standard deviation $\sigma(t)$, % or intensity
\begin{equation}
I_{\mathrm{syn}}(t) = C \left[ \mu(t) + \sigma(t) \eta(t) \right].
\label{I_syn_eq}
\end{equation}
% 
%INCLUDE IN METHODS: ...as typically observed in vivo (Anderson2000, Destexhe2001, 2003).
%Indeed, if the amplitudes of individual synaptic inputs are small, the compound input current resulting from many synaptic inputs can be represented by a Gaussian stochastic process in time (Stein 1965, 1967; Tuckwell 1988). %which is a good description of the activity in many cortical areas (Anderson et al. 2000; Destexhe et al. 2001, 2003), although not all (DeWeese and Zador 2006). 
%Except for Fig.1 we consider a stationary situation in which the mean and SD of the compound input do not vary in time. The values of this mean and SD depend on the synaptic strengths and on the firing rates of presynaptic cells.
Eq.~\eqref{I_syn_eq} describes the total synaptic current received by $K_{\EE}$ excitatory and $K_{\II}$ inhibitory neurons, which produce instantaneous postsynaptic potentials (PSPs) $J_{\EE} > 0$ and $J_{\II} < 0$, respectively. For synaptic events (i.e. presynaptic spike times) generated by independent Poisson processes with rates $r_{\EE}(t)$ and $r_{\II}(t)$, the infinitesimal moments $\mu(t)$ and $\sigma(t)$ are expressed as 
\begin{align}
\mu(t) &= J_{\EE}K_{\EE}r_{\EE}(t) + J_{\II}K_{\II}r_{\II}(t), \label{mu_eq} \\
\sigma(t)^2 &= J_{\EE}^2 K_{\EE}r_{\EE}(t) + J_{\II}^2 K_{\II}r_{\II}(t),
\label{sigma_eq}
\end{align}
assuming large numbers $K_{\EE}$, $K_{\II}$ and small magnitudes of $J_{\EE}$, $J_{\II}$ \textcolor{black}{\citep{Tuckwell1988,Renart2004,Destexhe2012}}. %Ricciardi1977 
This diffusion approximation well describes the activity in many cortical areas %Anderson2000?, Destexhe2001, 
\textcolor{black}{\citep{Shadlen1998,Destexhe2003,Compte2003,Maimon2009}}. %, but not all [DeWeese2006?]. 
The parameter values were $J_{\EE} = 0.15$~mV, $J_{\II} = -0.45$~mV, $K_{\EE}=2000$, $K_{\II}=500$ and $r_{\EE}$, $r_{\II}$ were varied in $[0,\,50]$~Hz. In addition, we directly varied $\mu$ and $\sigma$ over a wide range of biologically plausible values. \\

\noindent \textbf{Membrane voltage distribution and spike rate} \\
In the following we describe how we obtain the distribution of the membrane voltage $p(V,t)$ and the instantaneous spike rate $r(t)$ of a single neuron at time $t$ for a large number $N$ of independent trials. Note that by trial we refer to a solution trajectory of the system of stochastic differential equations eq.~\eqref{membrane_eq}--\eqref{I_syn_eq} for a realization of $\eta(t)$.

First, to reduce computational demands and enable further analysis, we replace the adaptation current $w$ in eqs.~\eqref{ionic_currents_eq}--\eqref{w_current_eq} by its average over trials, $\bar{w}(t) := 1/N \sum_{i=1}^N w_i(t)$, where $i$ is the trial index \citep{Gigante2007MB}. Neglecting the variance of $w$ across trials is valid under the assumption that the dynamics of the adaptation current is substantially slower than that of the membrane voltage, which is supported by empirical observations \citep{Brown1980,Sanchez-Vives2000,Sanchez-Vives2000b,Stocker2004}. %\textcolor{black}{see also Fig.~\ref{fig0}A}.
%Adiabatic approx: take average of w ("over many indep. realizations of orig. system") - justified under the assumption that fluctuations in w are negligible - in turn justified by separation of timescales (for V, w) 
The instantaneous spike rate at time $t$ can be estimated by the average number of spikes in a small interval $[t,\, t+ \Delta t]$,
\begin{equation}
    r_{\Delta t}(t) := \frac{1}{N \Delta t} \sum_{i=1}^{N} 
    \int_{t}^{t+\Delta t} \sum_{k} \delta(s-t_i^k) ds,
\end{equation}
where \textcolor{black}{$\delta$ is the delta function and} $t_i^k$ denotes the $k$-th spike time in trial $i$. In the limit ${N \to \infty}$, ${\Delta t \to 0}$, the probability
density $p(V,t)$ obeys the Fokker-Planck equation \citep{Risken1996,Tuckwell1988,Renart2004},
\begin{equation}
\frac{\partial}{\partial t} p(V,t) + \frac{\partial}{\partial V} q(V,t) = 0,
\label{FP_eq}
\end{equation}
with probability flux $q(V,t)$ given by
\begin{equation}
q(V,t) := \left( \frac{I_{\mathrm{ion}}(V; \bar{w})}{C} + \mu(t) \right) p(V,t) - \frac{\sigma(t)^2}{2} \frac{\partial}{\partial V} p(V,t).
\label{p_flux_eq}
\end{equation}
$I_{\mathrm{ion}}(V; \bar{w})$ denotes the sum of ionic currents (cf. eq.~\eqref{ionic_currents_eq}) where $w$ is replaced by the average adaptation current $\bar{w}$ which evolves according to
\begin{equation}
  \tau_w \frac{d\bar{w}}{dt} = a(\langle V \rangle_{p(V,t)} -E_w) - 
  \bar{w} + \tau_w \,b\, r(t).
  \label{w_mean_eq}
\end{equation}
%
%\citep{Brunel2003,Gigante2007PRL}.
$\langle \cdot \rangle_p$ indicates the average with respect to the probability 
density $p$ 
\citep{Brunel2003,Gigante2007PRL}.
To account for the reset of the membrane voltage, the probability flux at $V_{\mathrm{s}}$ is re-injected at $V_{\mathrm{r}}$ after the refractory period has passed, i.e.,
\begin{align}
    \lim_{V \searrow V_{\mathrm{r}}} q(V,t) 
    - \lim_{V \nearrow V_{\mathrm{r}}} q(V,t) = q(V_{\mathrm{s}}, t-T_{\mathrm{ref}}).
    \label{reinj_cond}
\end{align}  
The boundary conditions for this system are reflecting for ${V \to -\infty}$ and absorbing for $V = V_{\mathrm{s}}$,
%
%\begin{align}
%    \lim_{V \to -\infty} q(V,t) &= 0, \label{refl_bc} \\
%    p(V_{\mathrm{s}},t) &= 0, \label{absorb_bc}
%\end{align} 
\begin{equation}
    \lim_{V \to -\infty} q(V,t) = 0, \qquad
    p(V_{\mathrm{s}},t) = 0, \label{boundary_cond}
\end{equation} 
and the (instantaneous) spike rate is obtained by the probability flux at $V_{\mathrm{s}}$,
\begin{equation}
    r(t) = q(V_{\mathrm{s}}, t).
    \label{rate_eq}
\end{equation}
\textcolor{black}{Note that $p(V,t)$ only reflects the proportion %or fraction 
of trials where the neuron is not refractory at time $t$, given by $P(t) = \int_{-\infty}^{V_{\mathrm{s}}} p(v,t) dv$ ($<1$ for $T_{\mathrm{ref}} > 0$ and $r(t)>0$). The total probability density that the membrane voltage is $V$ at time $t$ is given by $p(V,t) + p_{\mathrm{ref}}(V,t)$, with refractory density $p_{\mathrm{ref}}(V,t) = (1-P(t)) \, \delta (V-V_{\mathrm{r}})$.
Since $p(V,t)$ does not integrate to unity in general, the average %membrane voltage with respect to $p(V,t)$ 
in eq.~\eqref{w_mean_eq} is calculated as
$\langle V \rangle_{p(V,t)} = \int_{-\infty}^{V_{\mathrm{s}}} v p(v,t) dv / P(t)$.
The dynamics of the average adaptation current $\bar{w}(t)$ reflecting the non-refractory proportion of trials is well captured by eq.~\eqref{w_mean_eq} %considering the refractory or non-refractory proportion of trials, or both, % which also only reflects the proportion of trials $P(t)$
as long as $T_{\mathrm{ref}}$ is small compared to $\tau_w$. In this (physiologically plausible) case $\bar{w}(t)$ can be considered equal to the average adaptation current over the refractory proportion of trials.} \\

\noindent \textbf{Steady-state spike rate} \\
We consider the membrane voltage distribution of an aEIF neuron with noisy synaptic input, described by the equations~\eqref{FP_eq}--\eqref{rate_eq}, has reached its steady-state $p_{\infty}$. $p_{\infty}$ obeys
$\partial p_{\infty}(V) / \partial t = 0$ or equivalently,
\begin{equation}
\frac{\partial}{\partial V} q_{\infty}(V) = 0,
\label{FPss_eq}
\end{equation}
with steady-state probability flux $q_{\infty}$ given by
\begin{equation}
q_{\infty}(V) = \left( \frac{I_{\mathrm{ion}}(V; \bar{w})}{C} + \mu \right) p_{\infty}(V) - \frac{\sigma^2}{2} \frac{\partial}{\partial V} p_{\infty}(V),
\label{p_fluxss_eq}
\end{equation}
subject to the reset condition, 
\begin{equation}
\lim_{V \searrow V_{\mathrm{r}}} q_{\infty}(V) - 
\lim_{V \nearrow V_{\mathrm{r}}} q_{\infty}(V) = q_{\infty}(V_{\mathrm{s}}),
\label{re_inj_ss}
\end{equation}
and the boundary conditions, 
\begin{equation}
\lim_{V \to -\infty} q_{\infty}(V) = 0 \qquad
p_{\infty}(V_{\mathrm{s}}) = 0
\label{boundary_cond_ss}
\end{equation}
The steady-state spike rate is given by 
${r_{\infty} = q_{\infty}(V_{\mathrm{s}})}$ and the steady-state mean adaptation current reads \linebreak
${\bar{w}_{\infty} = a (\langle V \rangle_{\infty} -E_w) + \tau_w b r_{\infty}}$. 
%We assume that $p_{\infty}(V)$ tends quickly toward zero for $V \to -\infty$ to be integrable [Brunel2000]. %, i.e., ${\lim_{V \to -\infty} V^2 \, p_{\infty}(V) = 0}$. 
We multiply both sides of eq.~\eqref{FPss_eq} by $V$ and integrate over the interval $(-\infty, V_{\mathrm{s}}]$, assuming that $p_{\infty}(V)$ tends sufficiently quickly toward zero for $V \to -\infty$ \citep{Brunel2000,Brunel2003}, to obtain %after simple manipulations,
an equation which relates the steady-state spike rate and mean membrane voltage,
\begin{equation}
r_{\infty} = \frac{
\mu_a - g_{\mathrm{L}} \left[
\langle V \rangle_{\infty} \! -\! E_{\mathrm{L}} \! + \!
\Delta_{\mathrm{T}} \left\langle \exp 
\left(\frac{V-V_{\mathrm{T}}}{\Delta_{\mathrm{T}}}
\right) \right\rangle_{\!\infty}
\right]/C}{\Delta V + \tau_w b /C},
\label{rate_Vmean_eq}
\end{equation}
%
%which relates the steady-state spike rate and mean membrane voltage. 
where ${\mu_a := \mu - a(\langle V \rangle_{\infty} \! -\! E_w)/C}$, 
$\Delta V := V_{\mathrm{s}}-V_{\mathrm{r}}$ (here and in the following) and 
$\langle \cdot \rangle_{\infty}$ denotes the average with respect to the density $p_{\infty}(V)$. \textcolor{black}{%Because of the above assumption for $p_\infty(V)$ which is only fulfilled when the (total) drift is nonnegative, which is only guaranteed if the numerator in eq. 18 is nonnegative (but only for positive rates) -> therefore: 
The spike rate $r_{\infty}$ is given by eq.~\eqref{rate_Vmean_eq} only for nonnegative values of the numerator (i.e., $\mu_a - g_{\mathrm{L}}[ \dots ] / C \geq 0$); otherwise, $r_{\infty}$ is defined to be zero.}
%In eq.~\eqref{rate_Vmean_eq} we have used ..., and the conditions eqs.~\eqref{re_inj_ss}--\eqref{boundary_cond_ss}.
For simplicity, the refractory period $T_{\mathrm{ref}}$ is omitted here. Note, that the steady-state spike rate for $T_{\mathrm{ref}} \neq 0$ can be calculated as ${r_{\infty}/(1+r_{\infty}T_{\mathrm{ref}})}$.
We cannot express $p_{\infty}(V)$ explicitly and thus the expressions for the averages with respect to $p_{\infty}(V)$ in eq.~\eqref{rate_Vmean_eq} are not known. However, in the case $g_{\mathrm{L}}=0$, which simplifies the aEIF model to the aPIF model, an explicit expression for $\langle V \rangle_{\infty}$ can be derived. We multiply eq.~\eqref{FPss_eq} by $V^2$ and integrate over $[-\infty, V_{\mathrm{s}}]$ on both sides \textcolor{black}{(assuming again that $p_{\infty}(V)$ quickly tends to zero for $V \to -\infty$)} to obtain
%
%\begin{align}
%\langle V \rangle_{\infty} =& \frac{1}{2a} \left[ \mu C + a \left(E_w +
%\frac{V_{\mathrm{s}}+V_{\mathrm{r}}}{2} \right) \right. \nonumber \\
%&- \left. \sqrt{\left[ \mu C + a \left(E_w -
%\frac{V_{\mathrm{s}}+V_{\mathrm{r}}}{2} \right) \right]^2 \!+ 
%2aC \sigma^2 \left[ 1+ \frac{\tau_w b}{C(V_{\mathrm{s}}-V_{\mathrm{r}})} \right]} \right]
%\end{align}
%
\begin{align}
\langle V \rangle_{\infty} = \frac{1}{2a} \left[ A +
a \frac{V_{\mathrm{s}}+V_{\mathrm{r}}}{2}
- \sqrt{\left( A -
a \frac{V_{\mathrm{s}}+V_{\mathrm{r}}}{2} \right)^2 \!+ B} \right],
\label{V_mean_expr}
\end{align}
where $A = \mu C + a E_w$ and ${B = 2a \sigma^2 C [ 1+ \tau_w b / (C \Delta V )]}$. \\

\noindent \textcolor{black}{\textbf{I-O curve}} \\
\textcolor{black}{The I-O curve is specified by the spike rate as a function of input strength. Here we consider two types of I-O curves: a time-varying (adapting) I-O curve and the steady-state I-O curve. In particular, we obtain the adapting I-O curve as the instantaneous spike rate response %of a neuron 
to a sustained input step (with a small baseline input) as a function of step size. This curve changes (adapts) over time and it eventually converges to the steady-state I-O curve. 
As arguments of these (adapting and steady-state) I-O functions we consider presynaptic spike rates (Figs.~\ref{fig1}C, \ref{fig3}B and eq.~\eqref{PIF_net_rate}), input mean and standard deviation\footnote{\textcolor{black}{Note that because of two arguments we obtain a surface instead of a curve in this case.}} (Figs.~\ref{fig1}D, \ref{fig3}B and eq.~\eqref{aPIF_rate_Vmean}) and 
input mean for fixed values of input standard deviation (Fig.~\ref{fig7}A).} \\
%I-O curves are presented by the (adapting) spike rate as a function of presynaptic spike rates (Figs.~\ref{fig1}C and \ref{fig3}B), as a function of mean and standard deviation of the input current (Figs.~\ref{fig1}D, \ref{fig3}B and eq.~\eqref{PIF_net_rate}) and as a function of mean input current for fixed values of input current standard deviation (Fig.~\ref{fig7}A and eq.~\eqref{aPIF_rate_Vmean}). %Rationale: why these definitions?
  
\noindent \textbf{ISI distribution} \\
We calculate the ISI distribution for an aEIF neuron which has reached a steady-state spike rate $r_{\infty} := \lim_{t\to\infty} r(t)$ by solving the so-called first passage time problem \citep{Risken1996,Tuckwell1988}. %We proceed similarly as in \citep{Ostojic2011}.
% MAYBE provide math. expression for ISI dist. as for instant. rate above
Consider an initial condition where the neuron has just emitted a spike and the refractory period has passed. That is, the membrane voltage is at the reset value $V_{\mathrm{r}}$ and the adaptation current, which we have replaced by its trial average (see above), takes the value $\bar{w}_0$, where $\bar{w}_0$ will be determined self-consistently (see below). In each of $N$ (simultaneous) trials, we follow the dynamics of the neuron given by
${dV_i /dt = [I_{\mathrm{ion}}(V_i;\bar{w}) + I_{\mathrm{syn}}(t)]/C}$,
${d\bar{w} /dt = [a(1/N \sum_{i=1}^N V_i - E_w) - \bar{w}]/\tau_w}$,
%
%\begin{align}
%C \frac{dV_i}{dt} &= I_{\mathrm{ion}}(V_i;\bar{w}) + I_{\mathrm{syn}}(t), \\
%\tau_w \frac{d\bar{w}}{dt} &= a(\frac{1}{N} \sum_{i=1}^N V_i - E_w) - \bar{w},
%\end{align}
%
until its membrane voltage crosses the value $V_{\mathrm{s}}$ and record that spike time $T_i$. The set of times $T_i + T_{\mathrm{ref}}$ then gives the ISI distribution. Finally,  
we determine $\bar{w}_0$ by imposing that the mean ISI matches with the known steady-state spike rate, i.e., 
$1/N \sum_{i=1}^N T_i + T_{\mathrm{ref}} = r_{\infty}^{-1}$.
According to this calculation scheme, the ISI distribution can be obtained in the limit $N \to \infty$ by solving the Fokker-Planck system eqs.~\eqref{FP_eq}--\eqref{p_flux_eq} with mean adaptation current governed by
\begin{equation}
\tau_w \frac{d\bar{w}}{dt} = a(\langle V \rangle_{p(V,t)}  - E_w) - \bar{w},
\label{w_mean_ISI_eq}
\end{equation}
subject to the boundary conditions~\eqref{boundary_cond} %eqs.~\eqref{refl_bc}--\eqref{absorb_bc} 
and initial conditions 
$p(V,0) = \delta (V-V_{\mathrm{r}})$, 
$\bar{w}(0) = \bar{w}_0$. %, where $\bar{w}_0$ is obtained self-consistently (see below). %and $r(t) = 0$. 
% Furthermore, we need to set $r \equiv 0$ in eq.~\eqref{w_mean_eq} because ...  
%Note, that here the membrane voltage is not reset when it has reached $V_{\mathrm{s}}$, instead, we start a new trial. Therefore, the re-injection condition eq.~\eqref{reinj_cond} is omitted and $r(t) = 0$ in eq.~\eqref{w_mean_eq}. 
Note that the re-injection condition eq.~\eqref{reinj_cond} is omitted 
(see also the difference between eqs.~\eqref{w_mean_eq} and \eqref{w_mean_ISI_eq}) 
because here each trial $i$ ends once $V_i(t)$ crosses the value $V_{\mathrm{s}}$.
The ISI distribution is given by the probability flux at $V_{\mathrm{s}}$ \textcolor{black}{\citep{Tuckwell1988,Ostojic2011}}, taking into account the refractory period, 
\begin{equation}
p_{\mathrm{ISI}}(T) =
\begin{cases}
q(V_{\mathrm{s}}, T-T_{\mathrm{ref}}) & \text{for } T \geq T_{\mathrm{ref}} \\
0 & \text{for } T < T_{\mathrm{ref}}.
\end{cases}
\label{pISI_eq}
\end{equation}
%
%$p_{\mathrm{ISI}}(T) = q(V_{\mathrm{s}}, T-T_{\mathrm{ref}})$. 
Finally, %the initial adaptation current value $\bar{w}(0)$ 
$\bar{w}_0$ is determined self-consistently by requiring 
$\langle T \rangle_{p_{\mathrm{ISI}}} = r_{\infty}^{-1}$. 
The coefficient of variation (CV) of ISIs is then calculated as
\begin{equation}
\mathrm{CV} := \frac{\sqrt{ \langle T^2 \rangle_{p_{\mathrm{ISI}}} - 
\langle T \rangle_{p_{\mathrm{ISI}}}^2}}
{\langle T \rangle_{p_{\mathrm{ISI}}}}.
\label{cvISI_eq}
\end{equation}
An ISI CV value of $0$ indicates regular, clock-like spiking, whereas %($\mathrm{CV} = 0$ for clock-like spiking), 
for spike times generated by a Poisson process the ISI CV assumes a value of $1$. 
For a demonstration of the ISI calculation scheme described above see Fig.~\ref{fig0}. The results based on the Fokker-Planck equation and numerical simulations of the aEIF model with fluctuating input are presented for an increased subthreshold and spike-triggered adaptation current in separation.\\ 

\noindent \textbf{ISI CV for the aPIF model} \\
To calculate the ISI CV we need the first two ISI moments, cf. eq.~\eqref{cvISI_eq}. The mean ISI for the aPIF neuron model is simply calculated by the inverse of the steady-state spike rate, cf. eq.~\eqref{rate_Vmean_eq}, derived in the previous section,
\begin{equation}
\langle T \rangle_{p_{\mathrm{ISI}}} = r_{\infty}^{-1} = \frac{
\Delta V + \tau_w b /C
}{\mu_a},
\label{meanISI_eq}
\end{equation}
%
%where $\langle V \rangle_{\infty}$ is given by eq.~\eqref{V_mean_expr}.
%%assuming again $T_{\mathrm{ref}} = 0$ for simplicity.
%
%We approximate the second ISI moment for large $\tau_w$ and small $b$, by solving the first passage time problem for the Langevin equation ...  using the approach in [Urdapilleta 2011] ...
%%i.e. by solving the backward FP equation to obtain $p_{\mathrm{ISI}}$,
%%... with ... (eq. 5 from Urdapilleta2011) where $P_s(V_r, T) = ...$ is the survival probability and $bar{w}_0$ is determined self-consistently ...
%
\textcolor{black}{where we consider $\mu_a > 0$ (here and in the following).}
We approximate the second ISI moment by solving the first passage time problem for the Langevin equation 
\begin{equation}
\frac{dV}{dt} = \mu_a - \frac{\bar{w}_0}{C} \exp(-t/\tau_w)
+ \sigma \eta(t),
\label{aPIF_wob_eq}
\end{equation}
with initial membrane voltage $V_{\mathrm{r}}$ and boundary voltage $V_{\mathrm{s}}$. That is, we replace $\langle V \rangle_{p(V,t)}$ by its steady-state value $\langle V \rangle_{\infty}$ in eq.~\eqref{w_mean_ISI_eq}, %\textcolor{gray}{governing the dynamics of the mean adaptation current between spikes,} 
%${d\bar{w}/dt = [a(\langle V \rangle_{\infty} \! -\! E_w) - \bar{w}] / \tau_w}$,
which is justified by %the large adaptation time constant 
large $\tau_w$ (as already assumed).
The first passage time density (which is equivalent to $p_{\mathrm{ISI}}$) and the associated first two moments for this type of Langevin equation can be calculated as power series %series in powers of $\tau_w \bar{w}_0/C$
in the limit of small $\bar{w}_0$ \citep{Urdapilleta2011}. % maybe also: Urdapilleta2012
$\bar{w}_0$ is then determined self-consistently by imposing eq.~\eqref{meanISI_eq}. 
Here we approximate the second ISI moment by using only the most dominant term of the power series, which yields (the zeroth order approximation) \citep{Urdapilleta2011},
\begin{equation}
\langle T^2 \rangle_{p_{\mathrm{ISI}}} = \frac{
\sigma^2 \Delta V + 
\mu_a
\Delta V^2
}{\mu_a^3}.
\label{ISIsecmoment_eq}
\end{equation}
Including terms of higher order leads to a complicated expression for
$\langle T^2 \rangle_{p_{\mathrm{ISI}}}$ which has to be evaluated numerically. We additionally considered the first order term (not shown) and compared the results of both approximations (see Results).
Effectively, the approximation above, eq.~\eqref{ISIsecmoment_eq}, is valid for small levels of spike-triggered adaptation current and mean input, since $\bar{w}_0$ increases with $b$ and $\mu$. %and $\bar{w}_0=0$ if $b=0$.
Combining eqs.~\eqref{cvISI_eq},\eqref{meanISI_eq} and \eqref{ISIsecmoment_eq} the ISI CV reads
% MAYBE provide full equation here
%
\begin{equation}
\mathrm{CV} = \frac{
\sqrt{ \sigma^2 \Delta V / \mu_a -  
\tau_w^2 b^2/C^2 - 2\tau_w b \Delta V /C }
}{\Delta V + \tau_w b/C}.
\label{cvISI_aPIF}
\end{equation}

\noindent \textbf{Neuronal network} \\ %Network motif
To investigate the effects of recurrent (inhibitory) synaptic inputs on the neuronal response properties (spike rates and ISIs),
we consider a network instead of a single neuron, consisting of $N_{\EE}$ excitatory and $N_{\II}$ inhibitory aEIF neurons (with separate parameter sets). The two populations are recurrently coupled in the following way (see Fig.~\ref{fig3}A). % maybe include small figure showing the single neuron and network 
%Each neuron receives synaptic inputs from external %excitatory and inhibitory 
%neurons as described above.
Each excitatory neuron receives inputs from $K_{\EE\EE}^{\mathrm{ext}}$ external excitatory neurons which produce instantaneous PSPs of magnitude $J_{\EE\EE}^{\mathrm{ext}}$ with Poisson rate $r_{\EE\EE}^{\mathrm{ext}} (t)$. Analogously, each inhibitory neuron receives inputs from $K_{\II\EE}^{\mathrm{ext}}$ external excitatory neurons producing instantaneous PSPs of magnitude $J_{\II\EE}^{\mathrm{ext}}$ with Poisson rate $r_{\II\EE}^{\mathrm{ext}} (t)$.
In addition, each excitatory neuron receives inputs from $K_{\EE\II}^{\mathrm{rec}}$ randomly selected inhibitory neurons of the network %which produce postsynaptic potentials of magnitude 
with synaptic strength (i.e., instantaneous PSP magnitude) $J_{\EE\II}^{\mathrm{rec}}$ and each inhibitory neuron receives inputs from $K_{\II\EE}^{\mathrm{rec}}$ randomly selected excitatory neurons of the network %which produce postsynaptic potentials of magnitude 
with synaptic strength $J_{\II\EE}^{\mathrm{rec}}$. 
This network setup was chosen to examine the effects caused by recurrent inhibition and compare them to the effects produced by external inhibition for single neurons described above. To reduce the parameter space, recurrent connections within the two populations in the network were therefore omitted.
The total synaptic current for each neuron of the network can be described using eq.~\eqref{I_syn_eq}, where the parameters $\mu(t)$ and $\sigma(t)$ for excitatory neurons are given by
\begin{align}
\mu(t) &= J_{\EE\EE}^{\mathrm{ext}} 
K_{\EE\EE}^{\mathrm{ext}} r_{\EE\EE}^{\mathrm{ext}} (t) + J_{\EE\II}^{\mathrm{rec}} 
K_{\EE\II}^{\mathrm{rec}} r_{\II}^{\mathrm{pop}} (t), \label{muE_net_eq} \\
\sigma(t)^2 &= (J_{\EE\EE}^{\mathrm{ext}})^2 
K_{\EE\EE}^{\mathrm{ext}} r_{\EE\EE}^{\mathrm{ext}} (t) + 
\left( J_{\EE\II}^{\mathrm{rec}} \right)^2 K_{\EE\II}^{\mathrm{rec}} 
r_{\II}^{\mathrm{pop}}(t)
\label{sigmaE_net_eq}
\end{align}
and for inhibitory neurons,
\begin{align}
\mu(t) &= J_{\II\EE}^{\mathrm{ext}} 
K_{\II\EE}^{\mathrm{ext}} r_{\II\EE}^{\mathrm{ext}} (t) + J_{\II\EE}^{\mathrm{rec}} 
K_{\II\EE}^{\mathrm{rec}} r_{\EE}^{\mathrm{pop}} (t), \label{muI_net_eq} \\
\sigma(t)^2 &= (J_{\II\EE}^{\mathrm{ext}})^2 
K_{\II\EE}^{\mathrm{ext}} r_{\II\EE}^{\mathrm{ext}} (t) + 
\left( J_{\II\EE}^{\mathrm{rec}} \right)^2 K_{\II\EE}^{\mathrm{rec}} 
r_{\EE}^{\mathrm{pop}}(t)
\label{sigmaI_net_eq}
\end{align}
\citep{Brunel2000,Augustin2013}. %and Poisson statistics for the spike times with constant rates on small time intervals 
%$\mu_{\EE}^{\mathrm{ext}}(t)$, $\sigma_{\EE}^{\mathrm{ext}}(t)$ and $\mu_{\II}^{\mathrm{ext}}(t)$, $\sigma_{\II}^{\mathrm{ext}}(t)$ are the first two moments of the total external synaptic current for excitatory and inhibitory neurons, respectively, which can be expressed in terms of the spike rates of the external neurons, cf. eqs.~\eqref{mu_eq}--\eqref{sigma_eq}.
$r_{\EE}^{\mathrm{pop}} (t)$ and $r_{\II}^{\mathrm{pop}} (t)$ are the spike rates of the excitatory and inhibitory neurons of the network, respectively.
Here we consider large populations of %independent
neurons instead of a large number of %independent 
trials. In fact, averaging over a large number of trials in this setting is equivalent to averaging over large populations due to the random and sparse connectivity. In the limit ${N_{\EE}, N_{\II} \to \infty}$ we obtain a system two coupled Fokker-Planck equations, one for the excitatory population, described by eqs.~\eqref{FP_eq}--\eqref{rate_eq},\eqref{muE_net_eq},\eqref{sigmaE_net_eq}, and one for the inhibitory population, given by eqs.~\eqref{FP_eq}--\eqref{rate_eq},\eqref{muI_net_eq},\eqref{sigmaI_net_eq}. Note that $r(t)$ in eqs.~\eqref{w_mean_eq} and \eqref{rate_eq} is replaced by the spike rates of the excitatory and inhibitory populations, $r_{\EE}^{\mathrm{pop}} (t)$ and $r_{\II}^{\mathrm{pop}} (t)$, respectively.
We solve this system to obtain the steady-state spike rate for each population, 
$r_{\EE,\infty}^{\mathrm{pop}}$ and $r_{\II,\infty}^{\mathrm{pop}}$. Once these quantities are known, we calculate the ISI distribution, cf. eq.~\eqref{pISI_eq}, for the excitatory population (i.e. for any neuron of that population) as described above, using eqs.~\eqref{muE_net_eq}--\eqref{sigmaE_net_eq} for the (steady-state) moments of the synaptic current. The neuron model parameter values were as above for the single neuron, with $a=0.015$~mS/$\mathrm{cm}^2$, $b=0.1$~$\SI{}{\micro\ampere}/\mathrm{cm}^2$ for excitatory neurons and $a=b=0$ for inhibitory neurons, since adaptation was found to be weak in fast-spiking interneurons compared to pyramidal neurons \citep{LaCamera2006}. The network parameter values were
$J_{\EE\EE}^{\mathrm{ext}} = J_{\II\EE}^{\mathrm{ext}} = 0.15$~mV, 
$K_{\EE\EE}^{\mathrm{ext}} = K_{\II\EE}^{\mathrm{ext}} = 800$,  constant $r_{\EE\EE}^{\mathrm{ext}} \in [0, \, 80]$~Hz, 
$J_{\EE\II}^{\mathrm{rec}} \in [-0.75, \, -0.45]$~mV,
$K_{\EE\II}^{\mathrm{rec}} = 100$,
constant $r_{\II\EE}^{\mathrm{ext}} \in [6, \, 14]$~Hz,
$J_{\II\EE}^{\mathrm{rec}} \in [0.05, \, 0.2]$~mV and 
$K_{\II\EE}^{\mathrm{rec}} = 400$. \\
%
%We will further consider noisy recurrent inhibitory input in addition,
%so mu = mu_ext + mu_rec and sigma = sigma_ext + sigma_rec, and mu_rec and sigma_rec^2 are proportional to r. Assuming K_rec indep. inhibitory neurons spiking with a rate prop. to r [assuming PIF neurons: r_I = mu_I/(Vcut-Vr)] and all feed back to the neuron, mu_rec and sigma_rec are given by ...
%

\noindent \textbf{Numerical solution} \\
%In the simplified case ($g_{\mathrm{L}} = 0$) we treated the FP equations analytically. 
We treated the Fokker-Planck equations for the aPIF model analytically. In case of the aEIF model, we solved these equations forward in time using a first-order finite volume method on a non-uniform grid with $512$ grid points in the interval $[-200\mathrm{~mV}, \, V_{\mathrm{s}}]$ and the implicit Euler integration method with a time step of $0.1$~ms for the temporal domain. For more details on the numerical solution, we refer to \citep{Augustin2013}. \\
%Efficiency (faster than simulating many realizations/trials).

\noindent \textbf{Detailed conductance-based neuron model} \\
For validation purposes we used a biophysical Hodgkin-Huxley-type neuron model with different types of slow $\mathrm{K}^+$-currents.
%
%The model includes a leak current ($I_{\mathrm{L}}$), a spike-generating $\mathrm{Na}^+$-current ($I_{\mathrm{Na}}$), a delayed rectifier $\mathrm{K}^+$-current ($I_{\mathrm{K}}$), a high-threshold calcium current ($I_{\mathrm{Ca}}$) and a slow $\mathrm{K}^+$-current ($I_{\mathrm{K,slow}}$). The membrane potential $V$ obeys the current balance equation
%...
%where $C=$ is the membrane capacitance and $I(t)$ denotes the injected current. We separately considered three types of slow $\mathrm{K}^+$-current: a calcium-activated current ($I_{\mathrm{K,slow}} \equiv I_{\mathrm{KCa}}$) which is associated with the slow after-hyperpolarization following burst of spikes [refs], a sodium-activated current ($I_{\mathrm{K,slow}} \equiv I_{\mathrm{KNa}}$) [refs], and the voltage-dependent muscarine-sensitive (M-type) current ($I_{\mathrm{K,slow}} \equiv I_{\mathrm{M}}$) [refs]. 
%The leak current is given by 
%$I_{\mathrm{L}} = g_{\mathrm{L}} (V-E_{\mathrm{L}})$. All other ionic currents depend on the membrane potential in a non-linear way, as described by the Hodgkin-Huxley formalism (see Appendix). 
%
The membrane voltage $V$ of this neuron model obeys the current balance equation
\begin{equation}
C \frac{dV}{dt} = I - I_{\mathrm{L}} - I_{\mathrm{Na}} - 
I_{\mathrm{K}} - I_{\mathrm{Ca}} - I_{\mathrm{Ks}}, 
\label{HH_membrane_eq}
\end{equation}
where $C=1$~$\SI{}{\micro\farad}/\mathrm{cm}^2$ is the membrane capacitance and $I$ denotes the injected current. 
The ionic currents consist of a leak current, $I_{\mathrm{L}} = g_{\mathrm{L}} (V-E_{\mathrm{L}})$, a spike-generating $\mathrm{Na}^+$-current, $I_{\mathrm{Na}} = g_{\mathrm{Na}}(V) (V-E_{\mathrm{Na}})$, a delayed rectifier $\mathrm{K}^+$-current, $I_{\mathrm{K}} = g_{\mathrm{K}}(V) (V-E_{\mathrm{K}})$, a high-threshold $\mathrm{Ca}^{2+}$-current, $I_{\mathrm{Ca}} = g_{\mathrm{Ca}}(V) (V-E_{\mathrm{Ca}})$, and a slow $\mathrm{K}^+$-current $I_{\mathrm{Ks}}$. $g_{\mathrm{x}}$ denote the conductances of the respective ion channels and $E_{\mathrm{x}}$ are the reversal potentials.
We separately considered three types of slow $\mathrm{K}^+$-current: a $\mathrm{Ca}^{2+}$-activated current ($I_{\mathrm{Ks}} \equiv I_{\mathrm{KCa}}$) which is associated with the slow after-hyperpolarization following a burst of spikes \citep{Brown1983}, a $\mathrm{Na}^+$-activated current ($I_{\mathrm{Ks}} \equiv I_{\mathrm{KNa}}$) \citep{Schwindt1989}, and the voltage-dependent muscarine-sensitive (M-type) current ($I_{\mathrm{Ks}} \equiv I_{\mathrm{M}}$) \citep{Brown1980}. 
The leak current depends linearly on the membrane potential. All other ionic currents depend on $V$ in a non-linear way as described by the Hodgkin-Huxley formalism. We adopted the somatic model from \citep{Wang2003} and included the M-current with dynamics described (for the soma) by \citep{Mainen1996}.   
%We used the somatic current dynamics from [Wang03] for $I_\mathrm L$, $I_\mathrm{Na}$, $I_\mathrm K$, 
%$I_\mathrm{Ca}$, $I_\mathrm{KCa}$ and $I_\mathrm{KNa}$. Additionally we implemented the 
%current $I_\mathrm M$ as described for the soma in [Mainen1996]. 
The conductances underlying the currents $I_\mathrm{Na}$, $I_\mathrm K$, $I_\mathrm{Ca}$ 
and $I_\mathrm M$ 
are given by $g_\mathrm{Na} = \bar g_\mathrm{Na} m_\infty^3 h$, 
$g_\mathrm K = \bar g_\mathrm K n^4$, $g_\mathrm{Ca} = \bar g_\mathrm{Ca} s_\infty^2$ 
and $g_\mathrm M=\bar g_\mathrm M u$, respectively, with 
steady-state gating variables 
$m_\infty = \alpha_m/(\alpha_m + \beta_m)$, 
$\alpha_m = -0.4 (V+33)/(\exp(-(V+33)/10)-1)$, 
$\beta_m = 16 \exp(-(V+58)/12)$ and
$s_\infty = 1/[1+\exp(-(V+20)/9)]$. 
The dynamic gating variables $x \in h, n, u$ are 
governed by 
\begin{equation}
 \frac{dx}{dt} = \alpha_x (1-x) - \beta_x x,
\end{equation}
where $\alpha_h = 0.28 \exp(-(V+50)/10)$, 
$\beta_h = 4/[1+\exp(-(V+20)/10)]$, 
$\alpha_n = -0.04 (V+34)/[\exp(-(V+34)/10)-1]$, 
$\beta_n = 0.5 \exp(-(V+44)/25)$, 
$\alpha_u = 3.209 \cdot 10^{-4} (V+30)/[1-\exp(-(V+30)/9)]$ % in code: a_m_M
and $\beta_u= -3.209 \cdot 10^{-4} (V+30)/[1-\exp((V+30)/9)]$. % in code: b_m_M 
The channel opening and closing
rates $\alpha_x$ and $\beta_x$ 
are specified in $\mathrm{ms}^{-1}$ and 
the membrane voltage $V$ in the equations above is replaced by its value in mV. 
%The unit of the channel opening and closing
%rates $\alpha_x$ and $\beta_x$ 
%(including $x=m$) is $\mathrm{ms}^{-1}$ and the membrane voltage $V$ in the equations above is in mV.
The conductance for the $\mathrm{Ca}^{2+}$-activated slow $\mathrm{K}^+$-current $I_\mathrm{KCa}$ is given by
$g_\mathrm{KCa} = \bar g_\mathrm{KCa} [\mathrm{Ca}]/([\mathrm{Ca}]+\kappa)$, where 
the intracellular $\mathrm{Ca}^{2+}$-concentration $[\mathrm{Ca}]$ satisfies 
\begin{equation}
\frac{d[\mathrm{Ca}]}{dt} = -\alpha_\mathrm{Ca} I_\mathrm{Ca} - \frac{[\mathrm{Ca}]}{\tau_{\mathrm{Ca}}}
\end{equation}
with 
$\alpha_{\mathrm{Ca}} = 6.67 \cdot 10^{-4} \, \SI{}{\micro M \centi\metre^2/(\micro \ampere \, \milli\second)}$, $\tau_\mathrm{Ca}=\SI{240}{\milli\second}$
and
$\kappa = \SI{0.03}{\milli M}$.
The conductance for the $\mathrm{Na}^+$--activated slow $\mathrm{K}^+$-current $I_\mathrm{KNa}$ is described by 
$g_\mathrm{KNa} = \bar g_\mathrm{KNa} 0.37/(1 + (\varrho/[\mathrm{Na}])^{3.5})$ 
where $\varrho = \SI{38.7}{\milli M}$ and the intracellular $\mathrm{Na}^+$-concentration $[\mathrm{Na}]$ is governed by 
\begin{equation}
\frac{d[\mathrm{Na}]}{dt} = -\alpha_\mathrm{Na} - 3 \varphi \left( \frac{[\mathrm{Na}]^3}{[\mathrm{Na}]^3 + \vartheta^3} - \gamma \right)
\end{equation}
with 
$\alpha_\mathrm{Na} = \SI{0.3}{\micro M \centi\metre^2/(\micro\ampere \, \milli\second)}$, $\varphi=\SI{0.6}{\micro M/\milli\second}$, $\vartheta=\SI{15}{\milli M}$ and $\gamma=0.132$. 
We varied the peak conductances of the three slow $\mathrm{K}^+$-currents
$I_{\mathrm{KCa}}$, $I_{\mathrm{KNa}}$, $I_{\mathrm{M}}$ in 
the ranges 
$\bar{g}_{\mathrm{KCa}} \in [2,\, 8]$~$\SI{}{\milli\siemens/\centi\metre^2}$, 
$\bar{g}_{\mathrm{KNa}} \in [2,\, 8]$~$\SI{}{\milli\siemens/\centi\metre^2}$ \citep{Wang2003} and
$\bar{g}_{\mathrm{M}} \in [0.1,\, 0.4]$~$\SI{}{\milli\siemens/\centi\metre^2}$ \citep{Mainen1996}.
% $\bar{g}_{\mathrm{KCa}} \in [2,\, 8]$~mS/$\mathrm{cm}^2$, \\
% $\bar{g}_{\mathrm{KNa}} \in [2,\, 8]$~mS/$\mathrm{cm}^2$, \\
% $\bar{g}_{\mathrm{M}} \in [0.1,\, 0.4]$~mS/$\mathrm{cm}^2$.
The remaining parameter values were 
$C=1$~$\SI{}{\micro\farad/\centi\metre^2}$, 
$g_\mathrm{L}=\SI{0.1}{\milli\siemens/\centi\metre^2}$, 
$E_\mathrm{L}=\SI{-65}{\milli\volt}$, 
$E_\mathrm{Na}=\SI{55}{\milli\volt}$, 
$E_\mathrm{K}=\SI{-80}{\milli\volt}$, 
$E_\mathrm{Ca}=\SI{120}{\milli\volt}$
\citep{Wang2003}. 

%The slowly-varying $\mathrm{K}^+$-conductances (for the currents $I_{\mathrm{KCa}}$, $I_{\mathrm{KNa}}$ and $I_{\mathrm{M}}$) depend on $V$ either directly (for $I_{\mathrm{M}}$) or indirectly (for $I_{\mathrm{KCa}}$ and $I_{\mathrm{KNa}}$) via the intracellular calcium and sodium concentrations, respectively. 
The differences of the slow $\mathrm{K}^+$-currents ($I_{\mathrm{KCa}}$, $I_{\mathrm{KNa}}$ and $I_{\mathrm{M}}$) is effectively expressed by their steady-state voltage dependence and time constants. Therefore, we further
considered a range of biologically plausible steady-state conductance-voltage relationships and timescales using the generic description of a slow $\mathrm{K}^+$-current, $I_{\mathrm{Ks}} = \bar{g}_{\mathrm{Ks}} \,\omega(V) (V-E_{\mathrm{K}})$, with peak conductance $\bar{g}_{\mathrm{Ks}}$ and gating variable $\omega(V)$ given by
\begin{equation}
\tau_{\omega} \frac{d\omega}{dt} = \omega_{\infty}(V) - \omega , 
\label{HH_adapt_gating_eq}
\end{equation}
where $\omega_{\infty}(V) = 1/ [1 + \mathrm{exp}(-(V-\alpha)/ \beta)] $. The shape of the steady-state curve $\omega_{\infty}(V)$ was changed by the parameters $\alpha \in [-40, \, -10]$~mV (half-activation voltage), $\beta \in [6, \, 12]$~mV (inverse steepness) and the time constant $\tau_{\omega}$ was varied in $[100, \, 300]$~ms. %The peak conductance was $\bar{g}_{\mathrm{Ks}} = 1$~$\mathrm{mS}/\mathrm{cm}^2$.
The model equations were solved using a second order Runge-Kutta integration method with a time step of $10$~$\SI{}{\micro\second}$. 

To examine the effects of slow $\mathrm{K}^+$-currents on the I-O curve and ISI variability for noisy input, we additionally considered the synaptic current described by eq.~\eqref{I_syn_eq} for the detailed neuron model, i.e., we used $I \equiv I_{\mathrm{syn}}$ in eq.~\eqref{HH_membrane_eq}.
\\

\noindent \textbf{Subthreshold and spike-triggered components of biophysical slow $\mathrm{K}^+$-currents} \\
%We quantified the relationship between the adaptation components of the aEIF model (parameters $a$ and $b$ of the adaptation current) and their equivalents in detailed biophysical descriptions of slow $\mathrm{K}^+$-currents ... 
To assess how the relative levels of subthreshold adaptation conductance (parameter $a$) and spike-triggered adaptation current increments (parameter $b$) in the aEIF model reflect different types of slow $\mathrm{K}^+$-currents, we quantified their subthreshold and spike-triggered components using the detailed conductance-based neuron model. 
% and applied a simple fitting procedure outlined below.
First, we fit the steady-state adaptation current $w_{\infty} = a (V-E_w)$ from the aEIF model to the respective $\mathrm{K}^+$-current $I_{\mathrm{Ks}}$ of the Hodgkin-Huxley-type model in steady-state over a range of subthreshold values for the membrane voltage, $V \in [-70, \, -60]$~mV.
%\textcolor{gray}{using the method of least squares}. 
Thereby we obtained an estimate for $a$. In the second step, we measured the absolute and relative change of $I_{\mathrm{Ks}}$ elicited by one spike. This was done by injecting a slowly increasing current ramp into the detailed model neuron and measuring $I_{\mathrm{Ks}}$ just before and after the first spike that occurred. Specifically, the absolute change of current caused by a spike was given by $\Delta I_{\mathrm{Ks}} := I_{\mathrm{Ks}}(t_{\mathrm{s}}^{\mathrm{post}}) - I_{\mathrm{Ks}}(t_{\mathrm{s}}^{\mathrm{pre}})$, where the time points $t_{\mathrm{s}}^{\mathrm{pre}}$ and $t_{\mathrm{s}}^{\mathrm{post}}$ were defined by the times at which the membrane potential crosses a value close to threshold (we chose $-50$~mV) during the upswing and downswing of the spike, respectively. $\Delta I_{\mathrm{Ks}}$ provides an estimate for $b$.
The relative change of $\mathrm{K}^+$-current was $\Delta I_{\mathrm{Ks}}^{\mathrm{rel}} := \Delta I_{\mathrm{Ks}}/I_{\mathrm{Ks}}(t_{\mathrm{s}}^{\mathrm{pre}})$. 
Here we only fitted the parameters $a$ and $b$ of the aEIF model. For an alternative fitting procedure which comprises all model parameters, we refer to \citep{Brette2005}.  

%Synaptic current: Gaussian process / OU process (Prescott2008) / OU processes for conductances ge and gi (Destexhe2001), details in Appendix \\

%% file: section_results.tex
\section*{Results}

\noindent \textbf{Spike rate adaptation, gain and threshold modulation in single neurons} \\
We first examine the responses of single aEIF neurons with and without an adaptation current, receiving inputs from stochastically spiking presynaptic excitatory and inhibitory neurons. The compound effect of the individual synaptic inputs is represented by an ongoing fluctuating input current whose mean and standard deviation depend on the synaptic strengths and spike rates of the presynaptic cells (cf. eqs.~\eqref{I_syn_eq}--\eqref{sigma_eq} in Materials and Methods and Fig.~\ref{fig1}A).
%Here we examine the response properties of a neuron without an adaptation current, and with an adaptation current either driven by subthreshold membrane depolarization (through parameter $a$) or by spikes (parameter $b$). 
The neurons naturally respond to a sudden increase in spike rate of the presynaptic neurons \textcolor{black}{(an input step)} with an abrupt increase in spike rate and mean membrane voltage, see Fig.~\ref{fig1}B. Without an adaptation current, both quantities remain unchanged after that increase. % whereas otherwise, a slow build-up of adaptation current leads to a gradual decrease ...
%In Fig.~\ref{fig1}B we show how a sudden increase in spike rate of the presynaptic neurons changes the spike rate and mean membrane voltage of a neuron without adaptation and with an adaptation current which is either driven by subthreshold membrane voltage depolarization (through conductance $a$) or by spikes (through current increment $b$).
%In all three cases the neuron responds with an abrupt increase in spike rate and mean membrane voltage. Without an adaptation current, both quantities remain unchanged after that increase, while the presence of an adaptation current causes a gradual decrease thereof. 
In case of a purely subthreshold adaptation current ($a>0$, $b=0$ in the aEIF model) which is present already in absence of spiking, the rapid increase of mean membrane voltage causes the mean adaptation current to build up slowly, which in turn leads to a gradual decrease in spike rate and mean membrane voltage. Note that the mean membrane voltage is decreased in the absence of spiking (before the increase of input) compared to the neuron without adaptation. In case of a purely spike-triggered adaptation current ($a=0$, $b>0$ in the aEIF model), the sudden increase in spike rate leads to an increase of mean adaptation current, which again causes the spike rate and mean membrane voltage to decrease gradually. %The build-up of adaptation current and concurrent decrease of spike rate occurs slightly faster compared to the case of subthreshold adaptation.

The adapting I-O curve of neurons with and without an adaptation current, \textcolor{black}{that is, the time-varying spike rate response to a step in presynaptic spike rates as a function of the step size,} is shown in Fig.~\ref{fig1}C. Interestingly, the two types of adaptation current affect the spike rate response in different ways. A subthreshold adaptation current shifts the I-O curve subtractively and thus increases the threshold for spiking. In addition, it decreases the response gain for low (output) spike rates. If the adaptation current is driven by spikes on the other hand, the I-O curve changes divisively, that is, the response gain is reduced over the whole range of spike rate values but the response threshold remains unchanged. 
\textcolor{black}{It can be recognized that for a given type of adaptation current the adapting I-O curve evaluated shortly after the input steps and the steady-state I-O curve are changed qualitatively in the same way. Thus, for the following parameter exploration and analytical derivation we focus on (changes of) the steady-state I-O relationship.}
  
We next explore the effects of an adaptation current on the steady-state spike rate for a wide range of input statistics, that is, different values of the mean $\mu$ and the standard deviation $\sigma$ of the fluctuating total synaptic input, see Fig.~\ref{fig1}D.
If excitatory and inhibitory inputs are approximately balanced, the standard deviation $\sigma$ of the compound input is large compared to its mean $\mu$.
The spike rate increases with an increase of either $\mu$ or $\sigma$, or both. 
A subthreshold adaptation current increases the threshold for spiking in terms of $\mu$ as well as $\sigma$. 
%In addition, when $\sigma$ is large the gain of the spike rate as a function of $\mu$ is slightly decreased, particularly ... % (as can be seen in the horizontal spacing of the contour lines). 
%For small values of $\sigma$ however, that gain is unaffected.
A spike-triggered adaptation current however does not change the threshold for spiking but reduces the gain of the spike rate as a function of $\mu$ or $\sigma$. %\textcolor{gray}{or any monotonously increasing function of $\mu$ and $\sigma$}. 
Thus, the differential effects of both types of adaptation current are robust across different input configurations. %, each of which characterizes a curve in the ($\mu,\sigma$) plane as the input increases.
Note that the I-O curve as a function of mean input $\mu$ changes additively for increased levels of standard deviation $\sigma$ while its slope (i.e., gain) decreases, particularly for small values of $\mu$. This can be recognized by the contour lines in Fig.~\ref{fig1}D and is most prominent for increased subthreshold adaptation. Consequently, this type of adaptation current increases the sensitivity of the steady-state spike rate to noise intensity for low spike rates. 

In order to analytically demonstrate the differential effects of subthreshold and spike-triggered adaptation currents on the (steady-state) I-O curve, we consider the aPIF neuron model, %instead of the aEIF model
which is obtained by neglecting the leak conductance ($g_{\mathrm{L}}=0$) in the aEIF model. This allows to derive an explicit expression for the steady-state spike rate, % $r_\infty$. The reduced neuron model is the so-called perfect IF model which includes an adaptation current %with subthreshold and spike-triggered components 
%(aPIF model). 
%It is calculated as
%
\begin{equation}
r_{\infty} = \frac{
\mu - a(\langle V \rangle_{\infty} \! -\! E_w)/C
}{\Delta V + \tau_w b /C},
\label{aPIF_rate_Vmean}
\end{equation}
where the mean membrane voltage $\langle V \rangle_{\infty}$ with respect to the steady-state distribution $p_{\infty}(V)$ is given by eq.~\eqref{V_mean_expr} and $\Delta V := V_{\mathrm{s}}-V_{\mathrm{r}}$ is the difference between spike and reset voltage;
\textcolor{black}{$r_{\infty} = 0$ for $\mu < a(\langle V \rangle_{\infty} \! -\! E_w)/C$ (see Materials and Methods).}
Equation~\eqref{aPIF_rate_Vmean} mathematically demonstrates the subtractive component of the effect a subthreshold adaptation current ($a>0$) produces when the mean membrane voltage is larger than the reversal potential $E_w$ of the ($\mathrm{K}^+$) adaptation current. Taking the derivative of eq.~\eqref{aPIF_rate_Vmean} with respect to $\mu$ further reveals that an increase of $a$ reduces the gain when the input fluctuations ($\sigma$) are large compared to the mean ($\mu$). A spike-triggered adaptation current ($b>0$) however produces a purely divisive effect which can be pronounced even for small current increments $b$ if the adaptation timescale $\tau_w$ is large. \\

\noindent \textbf{Differential effects of adaptation currents on spiking variability} \\
We next investigate how adaptation currents affect ISIs for different input statistics. For that reason we calculate the distribution of times at which the membrane voltage of an aEIF neuron crosses the threshold $V_{\mathrm{s}}$ for the first time, which is equivalent to the distribution of ISIs (see Materials and Methods). 
%In the mathematical literature, this is a well-studied quantity called the first-passage time distribution. Instead of directly studying the first passage time distribution, here we adopt a complementary approach and focus on the membrane potential distribution p(V,t), which directly determines the ISI distribution (see Methods).
These ISI distributions are shown in Fig.~\ref{fig2}A for neurons with different levels of subthreshold or spike-triggered adaptation and a given input. An increase of either type of adaptation current (via parameters $a$ and $b$) naturally increases the mean ISI. Interestingly, while subthreshold adaptation leads to ISI distributions with long tails, spike-triggered adaptation causes ISI distributions with bulky shapes. These differential effects on the shape of the ISI distribution lead to opposite changes of the coefficient of variation (CV, cf. eq.~\eqref{cvISI_eq}) which quantifies the variability of ISIs. An increase of subthreshold adaptation curent produces an increase of CV, whereas an increase of spike-triggered adaptation current leads to a decreased ISI variability.
How these effects on the CV of ISIs depend on the statistics ($\mu$ and $\sigma$) of the fluctuating input is shown in Fig.~\ref{fig2}B,C. With or without an adaptation current, if the mean $\mu$ is large, that is, far above threshold, and the standard deviation $\sigma$ is comparatively small, the neuronal dynamics is close to deterministic and the firing is almost periodic, hence the CV is small. In contrast, if $\mu$ is close to the threshold and $\sigma$ is large (enough), the ISI distribution will be broad as indicated by the large CV. 
%Fig.~\ref{fig2}C showing the change of ISI variability caused by an increase of $a$ or $b$ reveals that 
A subthreshold adaptation current either leads to an increased CV or leaves the ISI variability unchanged. In case of a spike-triggered adaptation current the effect on the CV depends on the input statistics. This type of adaptation current causes a decrease of the high ISI variability in the region (of the $\mu , \sigma$-plane) where the mean input $\mu$ is small, and an increase of the low ISI variability for larger values of $\mu$. %Thus, spike-triggered adaptation seems to have a normalizing effect on ISI variability. % MAYBE REPHRASE THIS

We analytically derived an approximation of the ISI CV 
for the aPIF model, which emphasizes the opposite effects of the two types of adaptation current. It is obtained as
\begin{equation}
\mathrm{CV} = \frac{
\sqrt{ \sigma^2 \Delta V / \mu_a -  
\tau_w^2 b^2/C^2 - 2\tau_w b \Delta V /C }
}{\Delta V + \tau_w b/C}
\label{cvISI_aPIF_results}
\end{equation}
\textcolor{black}{(same as eq.~\eqref{cvISI_aPIF})}, where ${\mu_a := \mu - a[\langle V \rangle_{\infty} \! -\! E_w]/C}$ %, $\Delta V := V_{\mathrm{s}}-V_{\mathrm{r}}$ 
is the effective mean input \textcolor{black}{which is again assumed to be positive} and takes into account the counteracting subthreshold adaptation current. The steady-state mean membrane voltage $\langle V \rangle_{\infty}$ is given by eq.~\eqref{V_mean_expr} (see Materials and Methods). Equation~\eqref{cvISI_aPIF_results} mathematically demonstrates that an increase of subthreshold adaptation curent ($a>0$) causes an increase of CV as long as $\langle V \rangle_{\infty}$ is larger than $E_w$, that is, the mean membrane voltage is not too hyperpolarized. An increase of spike-triggered adaptation current ($b>0$) on the other hand leads to a reduction of ISI variability. Note that this approximation is only valid for small values of mean input ($\mu$) and adaptation current increment ($b$). %\textcolor{black}{(and positive values of $\mu_a$)}. 
It does not account for the increase of CV caused by spike-triggered adaptation for large levels $\mu$, cf. Fig.~\ref{fig2}C. 
%The approximation can be refined (see Appendix), which results in a complicated expression for the CV that requires numerical evaluation. This refined description however captures the differential effects of $b$ qualitatively (not shown). 
Both (input dependent) effects of spike-triggered adaptation on the ISI variability can be captured by a refined approximation of the CV compared to eq.~\eqref{cvISI_aPIF_results} (not shown, see Materials and Methods for an outline), which requires numerical evaluation.
\\

\noindent \textbf{Differential effects of synaptic inhibition on I-O curves} \\
Here we examine how synaptic input received from a population of inhibitory neurons affect gain and threshold of spiking. % the I-O curve. 
We consider that the neuron we monitor %``record'' from 
belongs to a population of excitatory neurons which are recurrently coupled to neurons from an inhibitory population, as depicted in Fig.~\ref{fig3}A: 
Each neuron of the network receives excitatory synaptic input from external neurons and additional synaptic input from a number of neurons of the other population.
%In addition, each excitatory neuron receives synaptic input from a number of neurons of the inhibitory population and each inhibitory neuron receives synaptic input from a number of neurons of the excitatory population. 
The specific choice of the monitored excitatory neuron does not matter because of identical model parameters within each population and sparse random connectivity (see Materials and Methods).
Fig.~\ref{fig3}B shows how the \textcolor{black}{steady-state} I-O curve of excitatory neurons, i.e., the spike rate $r_{\EE,\infty}^{\mathrm{pop}}$ as a function of the external (input) spike rate $r_{\EE\EE}^{\mathrm{ext}}$, is changed by external excitation to the inhibitory neurons (via $r_{\II\EE}^{\mathrm{ext}}$) and by the strengths of the recurrent excitatory and inhibitory synapses ($J_{\II\EE}^{\mathrm{rec}}$ and $J_{\EE\II}^{\mathrm{rec}}$), respectively.
%We individually vary the excitatory drive $r_{\II}^{\mathrm{ext}}$ to the inhibitory neurons and the strengths of the recurrent excitatory and inhibitory synapses, $J_{\EE}^{\mathrm{rec}}$ and $J_{\II}^{\mathrm{rec}}$, respectively and measure the I-O curves of the excitatory neurons. 
An increase of external excitation to the inhibitory population (via $r_{\II\EE}^{\mathrm{ext}}$) changes the I-O curve subtractively, thus increasing the response threshold, while an increase of recurrent excitation to the inhibitory neurons (via $J_{\II\EE}^{\mathrm{rec}}$) has a purely divisive effect, that is, the gain is reduced. On the other hand, an increase of recurrent inhibition to the excitatory neurons (via $J_{\EE\II}^{\mathrm{rec}}$) affects the I-O curve in both ways. 
%That is, the spiking threshold of excitatory neurons increases while the gain of their spike rate response decreases.
%Inhibitory synaptic input reduces the mean total synaptic input $\mu$ and increases its standard deviation $\sigma$ for the target neuron (population), see Fig.~\ref{fig3}B,bottom.  
  
We demonstrate these effects analytically for a network of perfect integrate-and-fire (PIF) model neurons (instead of aEIF neurons). That is, we disregard the adaptation current here for simplicity ($a=b=0$), since it does not change the results qualitatively. An explicit expression for the steady-state spike rate of the excitatory neurons, $r_{\EE,\infty}^{\mathrm{pop}}$, %(of an excitatory neuron) 
can be derived using eq.~\eqref{aPIF_rate_Vmean} for all the neurons in the network with mean input $\mu$ given by eq.~\eqref{muE_net_eq} for excitatory neurons and by eq.~\eqref{muI_net_eq} for inhibitory neurons. We solve for $r_{\EE,\infty}^{\mathrm{pop}}$ self-consistently to obtain,
\begin{equation}
r_{\EE,\infty}^{\mathrm{pop}} = \frac{
J_{\EE\EE}^{\mathrm{ext}} 
K_{\EE\EE}^{\mathrm{ext}} r_{\EE\EE}^{\mathrm{ext}} \Delta V + 
J_{\EE\II}^{\mathrm{rec}} K_{\EE\II}^{\mathrm{rec}} J_{\II\EE}^{\mathrm{ext}} 
K_{\II\EE}^{\mathrm{ext}} r_{\II\EE}^{\mathrm{ext}}
}{\Delta V^2 - J_{\II\EE}^{\mathrm{rec}} K_{\II\EE}^{\mathrm{rec}} J_{\EE\II}^{\mathrm{rec}} K_{\EE\II}^{\mathrm{rec}}}.
\label{PIF_net_rate}
\end{equation}
The equation above states that $r_{\EE,\infty}^{\mathrm{pop}}$ is directly proportional to the strength of external excitation to the excitatory population, 
%($r_{\EE,\infty}^{\mathrm{pop}} \propto J_{\EE\EE}^{\mathrm{ext}} K_{\EE\EE}^{\mathrm{ext}} r_{\EE\EE}^{\mathrm{ext}}$), 
negatively proportional to the strength of external excitation to the inhibitory population 
%($r_{\EE,\infty}^{\mathrm{pop}} \propto -J_{\II\EE}^{\mathrm{ext}} K_{\II\EE}^{\mathrm{ext}} r_{\II\EE}^{\mathrm{ext}}$, 
(since $J_{\EE\II}^{\mathrm{rec}}  < 0$) and inversely proportional to the strength of recurrent excitation, 
%($r_{\EE,\infty}^{\mathrm{pop}} \propto 1/(J_{\II\EE}^{\mathrm{rec}} K_{\II\EE}^{\mathrm{rec}})$).
where all proportionalities include an offset.
Eq.~\eqref{PIF_net_rate} clearly shows that the effect of external excitation to the inhibitory population is purely subtractive (since $J_{\EE\II}^{\mathrm{rec}}  < 0$), the effect of %(the strength of) 
recurrent excitation (to the inhibitory population) is purely divisive, and the effect of %(the strength of) 
recurrent inhibition (to the excitatory population) includes both components.
For comparison, consider a single (non-adapting) PIF neuron receiving (external) excitatory and inhibitory input. 
Using eq.~\eqref{aPIF_rate_Vmean} with mean input $\mu$ given by eq.~\eqref{mu_eq}, the steady-state spike rate of this neuron reads ${r_{\infty} = (J_{\EE}K_{\EE}r_{\EE} + J_{\II}K_{\II}r_{\II}) / \Delta V}$.
Thus, an increase of external inhibition affects the I-O curve of an excitatory neuron in the same way (subtractively) as an increase of external excitation to the inhibitory population within a recurrent network as described above.
%%or:
%It should be pointed out that an increase of external excitatory input to the inhibitory population affects the I-O curve in the same way (subtractively) as an increase of inhibition received from external neurons. This can be recognized in eq.~\eqref{aPIF_rate_Vmean} with $\mu$ given by eq.~\eqref{mu_eq}, %(and $a=b=0$ for clarity), 
%which yields
%${r_{\infty} = (J_{\EE}K_{\EE}r_{\EE} + J_{\II}K_{\II}r_{\II}) / \Delta V}$
%for a perfect IF neuron. 
\\

\noindent \textbf{Effects of synaptic inhibition on spiking variability} \\
We next investigate how inhibitory synaptic input changes the ISI variability of the neurons (from the excitatory population) in the network described above. %(Fig.~\ref{fig3}A). 
%Fig.~\ref{fig4}A shows ISI distributions for different values of $r_{\II\EE}^{\mathrm{ext}}$, $J_{\II\EE}^{\mathrm{rec}}$ and $J_{\EE\II}^{\mathrm{rec}}$. 
An increase of external excitation to the inhibitory neurons (via $r_{\II\EE}^{\mathrm{ext}}$), and the strengths of the recurrent synapses ($J_{\II\EE}^{\mathrm{rec}}$ and $J_{\EE\II}^{\mathrm{rec}}$) individually, leads to an increase of the mean ISI and an increased tail of the ISI distribution, as shown in Fig.~\ref{fig4}A. Furthermore, an increase of $r_{\II\EE}^{\mathrm{ext}}$ or the magnitude of $J_{\II\EE}^{\mathrm{rec}}$ or $J_{\EE\II}^{\mathrm{rec}}$, each causes the coefficient of variation of ISIs ($\mathrm{CV}_{\EE}^{\mathrm{pop}}$) to increase, see Fig.~\ref{fig4}B. 
Thus, an increase of inhibition always leads to an increase of spiking variability.
An increase of external excitation to the excitatory neurons (via $r_{\EE\EE}^{\mathrm{ext}}$), on the other hand, leads to a decrease of $\mathrm{CV}_{\EE}^{\mathrm{pop}}$.

To demonstrate these effects analytically we derived $\mathrm{CV}_{\EE}^{\mathrm{pop}}$ for a network of PIF model neurons using eqs.~\eqref{cvISI_aPIF}--\eqref{sigmaE_net_eq}, where we obtained the steady-state spike rate of the inhibitory neurons, $r_{\II,\infty}^{\mathrm{pop}}$, analogously to $r_{\EE,\infty}^{\mathrm{pop}}$ (as described above). 
%
%... analytically using eq.~\eqref{cvISI_aPIF_results} with $\mu$ and $\sigma$ given by eqs.~\eqref{muE_net_eq} and \eqref{sigmaE_net_eq}, respectively, where the steady-state spike rate $r_{\II, \infty}^{\mathrm{pop}}$ (for eqs.~\eqref{muE_net_eq}, \eqref{sigmaE_net_eq}) is obtained analogously to $r_{\EE, \infty}^{\mathrm{pop}}$ as described above (cf. eq.~\eqref{PIF_net_rate}). The coefficient of variation is given by
%%
%\begin{equation}
%\mathrm{cv}(T) = \sqrt{ \frac{
%(\sigma_{\EE}^{\mathrm{ext}})^2 (\Delta \! V^2 - J_{\EE}^{\mathrm{rec}}
%K_{\EE}^{\mathrm{rec}} J_{\II}^{\mathrm{rec}} K_{\II}^{\mathrm{rec}} ) +  
%(J_{\II}^{\mathrm{rec}})^2 K_{\II}^{\mathrm{rec}} (\mu_{\II}^{\mathrm{ext}}
%\Delta \! V + J_{\EE}^{\mathrm{rec}} K_{\EE}^{\mathrm{rec}} \mu_{\EE}^{\mathrm{ext}})
%}
%{\Delta \! V^2 ( \mu_{\EE}^{\mathrm{ext}} \Delta \! V + J_{\II}^{\mathrm{rec}} K_{\II}^{\mathrm{rec}} \mu_{\II}^{\mathrm{ext}}) }}.
%\label{cvISI_PIF_net}
%\end{equation}
%%
Below, we express $\mathrm{CV}_{\EE}^{\mathrm{pop}}$ as a function of either $r_{\II\EE}^{\mathrm{ext}}$, $J_{\II\EE}^{\mathrm{rec}}$ or $J_{\EE\II}^{\mathrm{rec}}$, and lump together all other fixed parameters in a number of constants, 
%The CV as a function of $r_{\EE\II}^{\mathrm{ext}}$ takes the form 
%${cv = (c_1 r_{\EE\II}^{\mathrm{ext}} + c_2) / (c_3 - c_4 r_{\EE\II}^{\mathrm{ext}})}$, in case of only varying $J_{\II\EE}^{\mathrm{rec}}$ it reads ${cv = c_5 J_{\II\EE}^{\mathrm{rec}} + c_6}$ and it can be expressed as a function of $J_{\EE\II}^{\mathrm{rec}}$, ${cv = (c_7 (J_{\EE\II}^{\mathrm{rec}})^2 - c_8 J_{\EE\II}^{\mathrm{rec}}) / (c_9 + c_{10} J_{\EE\II}^{\mathrm{rec}})}$. 
%
\begin{equation}
\mathrm{CV}_{\EE}^{\mathrm{pop}} = 
\begin{cases}
(c_1 r_{\II\EE}^{\mathrm{ext}} + c_2) / (c_3 - c_4 r_{\II\EE}^{\mathrm{ext}}) \\
c_5 J_{\II\EE}^{\mathrm{rec}} + c_6 \\
(c_7 (J_{\EE\II}^{\mathrm{rec}})^2 - c_8 J_{\EE\II}^{\mathrm{rec}}) / (c_9 + c_{10} J_{\EE\II}^{\mathrm{rec}}).
\end{cases} \label{cvISI_PIF_net}
\end{equation}
The constants $c_1, \dots, c_{10}$ in eq.~\eqref{cvISI_PIF_net} are non-negative functions of the fixed parameters.
% , their precise values depend on the fixed parameter values. 
Clearly, an increase of $r_{\II\EE}^{\mathrm{ext}}$ or the magnitudes of $J_{\II\EE}^{\mathrm{rec}}$ and $J_{\EE\II}^{\mathrm{rec}}$ each produce an increase of $\mathrm{CV}_{\EE}^{\mathrm{pop}}$ (since $J_{\EE\II}^{\mathrm{rec}}  < 0$).
Considering a single PIF neuron receiving (external) excitatory and inhibitory input for comparison, we use eq.~\eqref{cvISI_aPIF_results} with mean $\mu$ and standard deviation $\sigma$ of the input given by eqs.~\eqref{mu_eq} and \eqref{sigma_eq}, respectively, %(and $a=b=0$ for simplicity), 
to express the CV as
\begin{equation}
\mathrm{CV} = \sqrt{
\frac{J_{\EE}^2 K_{\EE}r_{\EE} + J_{\II}^2 K_{\II}r_{\II}}
{\Delta V (J_{\EE}K_{\EE}r_{\EE} + J_{\II}K_{\II}r_{\II})}
}. \label{cvISI_PIF_net_simple}
\end{equation} 
\textcolor{black}{Note that eq.~\eqref{cvISI_PIF_net_simple} is only valid for positive mean input ($J_{\EE}K_{\EE}r_{\EE} + J_{\II}K_{\II}r_{\II} >0$).}
Again, ISI variability increases with inhibition.
The effect of inhibition on spiking variability can be understood intuitively as follows. Inhibitory synaptic input reduces the mean total synaptic input $\mu$ and increases its standard deviation $\sigma$ for the target neuron (population), which in turn causes an increase of ISI variability. \\

\noindent \textbf{Subthreshold and spike-triggered components of slow $\mathrm{K}^+$-currents} \\
%To determine which type of $\mathrm{K}^+$ current is represented by either a subthreshold voltage dependent or a spike dependent description, or both, we quantify the subthreshold and spike-triggered components of such currents in a biophysical HH type neuron model. 
Here we examine how the two types of an adaptation current in the aEIF model reflect different slow $\mathrm{K}^+$-currents in a detailed conductance-based neuron model. 
%... we quantify the subthreshold and spike-triggered components of slow $\mathrm{K}^+$-currents in a detailed conductance-based neuron model.
%%to determine which current is represented by either a subthreshold voltage dependent or a spike dependent description, or both.
%
%That is, we fit the adaptation current from the aEIF model to different $\mathrm{K}^+$-currents described biophysically using the Hodgkin-Huxley formalism (see Materials and Methods). 
First, we consider three prominent slow $\mathrm{K}^+$-currents: a $\mathrm{Ca}^{2+}$-activated after-hyperpolarization current ($I_{\mathrm{KCa}}$), a $\mathrm{Na}^+$-activated current ($I_{\mathrm{KNa}}$) and the voltage-dependent M-current ($I_{\mathrm{M}}$). 
Fig.~\ref{fig5}A shows how the conductances associated with these $\mathrm{K}^+$-currents depend on the membrane voltage in the steady state, compared to the steady-state spike-generating $\mathrm{Na}^+$-conductance. 
%Note that $I_{\mathrm{KCa}}$ and $I_{\mathrm{KNa}}$ depend on the ion-channel (gating) kinetics of $I_{\mathrm{Ca}}$ and $I_{\mathrm{Na}}$, respectively. 
%TODO: include I_KNa, (exclude K)
The threshold membrane voltage at which a spike is elicited in response to a slowly increasing input current is primarily determined by the conductance-voltage relationship for $\mathrm{Na}^+$. The threshold value lies in the interval where this curve has a positive slope (the precise value depends on the peak conductances of all currents and on the input). The curve $g_{\mathrm{Na},\infty}(V)$ thus indicates the subthreshold and suprathreshold membrane voltage ranges. 
In the subthreshold voltage range the conductance $g_{\mathrm{KCa},\infty}$ is almost zero, while the conductances $g_{\mathrm{KNa},\infty}$ and $g_{\mathrm{M},\infty}$ reach significant values close to the voltage threshold. 
%The conductance for $I_{\mathrm{KCa}}$ on the other hand is practically zero for subthreshold voltage values. 
Thus, the curves in Fig.~\ref{fig5}A indicate that $I_{\mathrm{KCa}}$ is activated by spikes, while $I_{\mathrm{M}}$ and particularly $I_{\mathrm{KNa}}$ can be increased in the absence of spiking.

The results of the fitting procedure in Fig.~\ref{fig5}B,C show the absolute and relative amounts of current triggered by a spike versus its subthreshold level quantified by the voltage independent conductance $a$.
$I_{\mathrm{KCa}}$ has a dominant spike-triggered component as expected, while $I_{\mathrm{KNa}}$ shows a very small increment caused by a spike compared to the subthreshold component. $I_{\mathrm{M}}$, on the other hand, shows significant levels of both components.
Note, however, that the amount of $I_{\mathrm{M}}$ elicited by a spike is smaller compared to the level of $I_{\mathrm{M}}$ that can be caused by subthreshold membrane depolarization without spiking (since $\Delta I_{\mathrm{Ks}}^{\mathrm{rel}} < 1$ for $I_{\mathrm{Ks}} \equiv I_{\mathrm{M}}$, see Fig.~\ref{fig5}C).
%Because of its dependence on intracellular calcium which enters the cell when the membrane voltage is strongly depolarized (Fig.5B) ...

We further considered a range of biologically plausible slow $\mathrm{K}^+$-currents. That is, we varied the steady-state conductance-voltage relationship for $\mathrm{K}^+$, $g_{\mathrm{Ks},\infty}(V)$, within a realistic range, as shown in Fig.~\ref{fig6}A, and quantified the subthreshold and spike-triggered components for each of these $\mathrm{K}^+$-currents, see Fig.~\ref{fig6}B,C.
%How the shape of the curve $g_{\mathrm{Ks},\infty}(V)$ determines the subthreshold level (quantified by $a$) as well as the spike-triggered absolute and relative current increment ($\Delta I_{\mathrm{Ks}}$ and $\Delta I_{\mathrm{Ks}}^{\mathrm{rel}}$, respectively) is shown in Fig.~\ref{fig6}B,C. 
The value of subthreshold conductance $a$ naturally increases with the fraction of $\mathrm{K}^+$-conductance present at subthreshold voltage values. For the quantification of spike-triggered current increments we also considered different $\mathrm{K}^+$ time constants $\tau_{\omega}$. The absolute value of current increment $\Delta I_{\mathrm{Ks}}$ decreases with increasing $\tau_{\omega}$ and changes only slightly with changes of the shape of the conductance-voltage curve $g_{\mathrm{Ks},\infty}(V)$ (via the parameters $\alpha$, $\beta$). However, the current increment caused by a spike relative to the amount of current already present in the absence of spiking ($\Delta I_{\mathrm{Ks}}^{\mathrm{rel}}$) is strongly determined by $g_{\mathrm{Ks}}(V)$. $\Delta I_{\mathrm{Ks}}^{\mathrm{rel}}$ increases with an increase of half-activation voltage (parameter $\alpha$), steepness (via parameter $\beta$) and with decreasing time constant ($\tau_{\omega}$). \\

\noindent \textbf{Effects of slow $\mathrm{K}^+$-currents %$I_{\mathrm{KCa}}$, $I_{\mathrm{KNa}}$ and $I_{\mathrm{M}}$ 
on I-O curve and ISI variability} \\
Here we examine how the different types of slow $\mathrm{K}^+$-current affect the I-O curve and spiking variability of uncoupled conductance-based %or: Hodgkin-Huxley-type 
model neurons subject to noisy inputs and compare the effects to those caused by subthreshold and spike-triggered adaptation in aEIF neurons.
Without a slow $\mathrm{K}^+$-current, the spike rate I-O curve does not change over time, see Fig.~\ref{fig7}A. An increase of $I_{\mathrm{KCa}}$ has a purely divisive effect on the I-O curve while an increase of $I_{\mathrm{M}}$ changes this curve in a mostly subtractive and slightly divisive way. For both types of slow $\mathrm{K}^+$-current the adapting spike rates reach their steady-state values in less than $500$~ms. These effects are consistent with our results based on the aEIF model, given that $I_{\mathrm{KCa}}$ predominantly depends on spikes and $I_{\mathrm{M}}$ includes both, subthrehold as well as spike-triggered, components (Fig.~\ref{fig5}B). 
In case of increased $I_{\mathrm{KNa}}$, on the other hand, the steady-state I-O curve is significantly altered in both ways (subtractively and divisively), and the spike rates adapts very slowly, that is, steady-state rates are reached after several seconds. At first sight, this seems contradictory to the effect predicted above for subthreshold adaptation, considering that the amount of $I_{\mathrm{KNa}}$ triggered by a spike is small compared to its subthreshold level. Since the timescale of $I_{\mathrm{KNa}}$ is very large (Fig.~\ref{fig7}A and \citep{Wang2003}) even a small spike-triggered component %(here represented by $\Delta I_{\mathrm{KNa}}$) 
leads to a significant divisive change of the steady-state I-O curve, cf. eq.~\eqref{aPIF_rate_Vmean}. This divisive effect is caused by $\mathrm{K}^+$-current building up slowly because of small current increments triggered repeatedly by repetitive spiking and very slow decay between spikes due to the large timescale of the current. 
%In this way, the spike (rate) dependence of $I_{\mathrm{KNa}}$ is expressed. 

Considering ISI variability, an increase of $I_{\mathrm{KCa}}$ reduces the CV for small values of mean input $\mu$ and increases the CV for larger values of $\mu$, see Fig.~\ref{fig7}B. An increase of each of the other slow $\mathrm{K}^+$-currents, $I_{\mathrm{KNa}}$ and $I_{\mathrm{M}}$, leads to an increase of ISI CV in general. These effects are consistent with those caused by subthreshold and spike-triggered adaptation currents in the aEIF model, considering the subthreshold and spike-triggered components of $I_{\mathrm{KCa}}$, $I_{\mathrm{KNa}}$ and $I_{\mathrm{M}}$, respectively (Fig.~\ref{fig5}).
Thus, the results from the detailed conductance-based neuron model are in agreement with the results based on the adaptive integrate-and-fire models presented above.

%%% MOTIVATE EACH (NEXT) STEP AS WELL AS POSSIBLE

%% file: section_discussion.tex
\section*{Discussion}

In this study, we have systematically examined how adaptation currents and synaptic inhibition modulate the threshold and gain of spiking as well as ISI variability %of model neurons 
in response to fluctuating inputs resulting from stochastic synaptic events.
Based on a simple neuron model with subthreshold and spike-triggered adaptation components we used analytical and numerical tools to describe spike rates and ISIs for a wide range of input statistics.
% ... perfect IF model that is amenable to analytical calculations
We then measured subthreshold and spike-triggered components of different types of slow $\mathrm{K}^+$-currents using detailed conductance-based model neurons and we validated our (analytical) results from the simple neuron model by numerical simulations of the detailed model. 

We have shown that a purely subthreshold voltage-dependent adaptation current increases the threshold for spiking and reduces the gain at low spike rates in the presence of input fluctuations. This type of current produces a long-tailed ISI distribution and thus leads to an increase of variability for a broad range of input statistics.
A spike-triggered adaptation current, on the other hand, causes a divisive
change of the I-O curve, thereby reducing the response gain but leaving the response threshold unaffected, irrespective of the input noise intensity. This type of current decreases the ISI CV for fluctuation-dominated inputs but increases the CV when the mean input is strong, i.e., it reduces the sensitivity of spiking variability to the mean input.
For comparison, an increase of external inhibition leads to a subtractive shift of the I-O curve while an increase of recurrent inhibition changes
it divisively. The ISI variability, however, is increased by
both types of synaptic inhibition.

We have further demonstrated that the $\mathrm{Ca}^{2+}$-activated after-hyperpolarization $\mathrm{K}^+$-current is effectively captured by a
simple description based on spike-triggered increments, while the muscarine-sensitive and $\mathrm{Na}^+$-activated $\mathrm{K}^+$-currents, respectively, have dominant subthreshold components. Despite its small spike-triggered component, the $\mathrm{Na}^+$-dependent $\mathrm{K}^+$-current also substantially affects the neuronal gain, due to its large timescale.
\\

%\textcolor{gray}{Note on importance of timescale when interpreting Fig.~5. e.g.: Therefore, timescale needs to be taken into account w.r.t. the spike-dependent adaptation component as far as the I-O curve is concerned, see eq. 22.}
%
%All important results first, then assumptions and theory, and then relation to exp. data (repeating our results)
%%Or: assumptions and theoretical studies first, then results plus exp. findings together, one by one

\noindent \textbf{Methodological aspects} \\
%Long history of rates for scalar IF models
Our approach involves the diffusion approximation and Fokker-Planck equation, both of which have been widely applied to analyze the spike rates of scalar IF type neurons in a noisy setting, see e.g. \citep{Amit1997,Brunel2000,Fourcaud-Trocme2003,Burkitt2006,Roxin2011}. Our assumption of separated timescales between slow adaptation and fast membrane voltage dynamics has also been frequently used in such a setting 
\citep{Brunel2003,LaCamera2004,Gigante2007PRL,Richardson2009,Augustin2013}. While most of these previous studies concentrated on spike rate dynamics, here we focused on asynchronous (non-oscillatory) activity. 
To examine ISI distributions we extended the method described \textcolor{black}{previously for scalar IF models, which is based on the first passage time problem \citep{Tuckwell1988,Ostojic2011}}, to the aEIF model, accounting for the dynamics of the adaptation current between spikes. Furthermore, we analytically derived an expression for the steady-state spike rate \textcolor{black}{(i.e., steady-state I-O relationship)} based on \citep{Brunel2003} and an approximation of the ISI CV using recent results from \citep{Urdapilleta2011} for the perfect IF model with two types of adaptation currents (aPIF model).
\textcolor{black}{The I-O functions we calculated can be used to relate (adaptive) spiking neuron models to linear-nonlinear cascade models, %(a generalized linear model class) 
which describe the instantaneous spike rate of a neuron by applying to the stimulus signal successively a linear temporal filter and a static nonlinear function \citep{Ostojic2011a}. Such cascade models have proven valuable for studying how sensory inputs are mapped to neuronal activity (see, e.g., \citep{Schwartz2006,Pillow2008}).}

It is worth noting that our approach further allows to easily calculate the power spectrum $\mathcal{P}$ and (normalized) autocorrelation function $\mathcal{A}$ of the neuronal spike train once the ISI distribution has been obtained, via the relation 
\begin{equation}
 \mathcal{P}(\omega) =  \hat{\mathcal{A}}(\omega) 
 = r_\infty \mathrm{Re} \! \left( \frac{1+\hat{p}_{\mathrm{ISI}}(\omega)}{1-\hat{p}_{\mathrm{ISI}}(\omega)} \right), 
 \label{eq_autocorr}
\end{equation}
where $\hat{\mathcal{A}}$ and $\hat{p}_{\mathrm{ISI}}$ denote the Fourier transforms of the autocorrelation function and ISI distribution, respectively, see \citep{Gerstner2002}. \textcolor{black}{Eq.~\eqref{eq_autocorr} strictly applies to memoryless (so-called renewal) stochastic processes and %a ``noisy'' neuron model with an adaptation mechanism usually violates this requirement. 
an adaptation mechanism usually leads to a violation of this requirement for a model neuron subject to fluctuating input. Here we have derived a renewal process $(V_i(t), \bar{w}(t))$ from the original non-renewal process $(V_i(t), w_i(t))$ by averaging the adaptation current and self-consistently determining its reset value (see section ISI distribution in Materials and Methods). An alternative approach that allows for the application of the above relationship eq.~\eqref{eq_autocorr} to adapting model neurons has recently been described in \citep{Naud2012}.}
% which have been neglected in a previous study analyzing ISI distributions of perfect IF model neurons with purely spike-triggered adaptation, see eq. (66) of \citep{Schwalger2010}.}
\\

\noindent \textbf{Modulation of spike rate threshold and gain} \\
Purely subtractive and divisive changes of the I-O curve by subthreshold and spike-triggered adaptation, respectively, have previously been shown for model neurons considering constant current inputs but neglecting input fluctuations \citep{Prescott2008,Ladenbauer2012}.
These theoretical results \textcolor{black}{describe the effects} shown in recent 
%have recently been supported by 
in-vitro experiments which involved blocking the low-threshold M current and a $\mathrm{Ca}^{2+}$-activated $\mathrm{K}^+$-current separately \citep{Deemyad2012} (Fig.~3); see also \citep{Alaburda2002} (Fig.~3), \citep{Smith2002} and \citep{Miles2005} (Fig.~1) for experimental evidence of either effect.
%\textcolor{gray}{It is worth noting that the spike-triggered component of the M current measured in \citep{Deemyad2012} seems weaker compared to the one estimated here for the HH based description of M current by \citep{Mainen1996}.} % OMIT perhaps
%
%... the leak current, which rather modulates the threshold [Shriki2003].   
%Fernandez2010: Experimental results, i.e., gain and threshold modulation by increasing membrane conductance. Which conductances are increased exactly is not clear: results suggest contribution of M and AHP conductances, but authors go for increase in sodium inactivation (in their modeling part)
%Patel2012: Gain modulation by low threshold potassium current (must have s.t. component, like M current here)
Here we have shown that a subthreshold adaptation current also causes a reduction of response gain (in addition to an increase of response threshold) when the fluctuations of the input are strong compared to its mean. On the other hand, a spike-triggered adaptation current decreases the response gain over the whole input range, irrespective of the level of input fluctuations. These results apply for adapting as well as the adapted (steady) states\footnote{\textcolor{black}{Note that in case of a very large adaptation timescale a (small) spike-triggered adaptation current has a negligible effect on the adapting I-O curve, evaluated shortly after the input steps, but a significant effect on the steady-state I-O curve (see Fig.~\ref{fig7}A).}}.
%It should be noted that %Counterintuitively,
When considering the onset I-O curve, i.e., the immediate response to a sudden increase of input, an increased level of spike-triggered adaptation current due to pre-adaptation has been shown to produce a rather subtractive change \citep{Benda2010}. This, however, does not contradict our results. On the contrary, either type of adaptation current (subthreshold or spike-triggered) naturally leads to a subtractive change of the onset I-O curve for neurons which are pre-adapted to an increased input (not shown).
 
Notably, when considering conductance based noisy synaptic inputs, an increase in balanced synaptic background activity can also reduce the spike rate gain \citep{Chance2002,Burkitt2003} %(Ly2009) 
and external inhibition can reduce the gain and increase the response treshold at the same time \citep{Mitchell2003}. 
This means, the response gain can change due to external inputs that are independent of the activity of the target neuron, which can be understood as follows. 
%These effects of noisy synaptic conductances can be decomposed into two separate effects [Chance2002] which in combination lead to the observed changes of the I-O curve. Both separate effects are consistent with our results.
%The increased synaptic background activity induced in-vitro [Chance2002, Mitchell2003] 
An increase of noisy (excitatory or inhibitory) synaptic conductance %or: in a noisy input scenario
leads to an increase of total membrane conductance, which causes a purely subtractive change of the I-O curve, %(see Fig.~3A in [Chance2002] and eq.~\eqref{rate_Vmean_eq} here) 
and an increase in synaptic current noise, which causes an additive change of the I-O curve and decreases its slope (particularly for small input strengths) \citep{Chance2002} (Fig.~3). Both effects combined lead to the observed change of response gain.
The two separate components are included in our results. An increase of membrane conductance (represented by $g_{\mathrm{L}}$ in the aEIF model) subtracts from the spike rate response, see eq.~\eqref{rate_Vmean_eq}, and the abovementioned effects of an increase of noise intensity $\sigma$ have been described in the section Results (see Fig.~\ref{fig1}D).

%... the immediate response to a sudden increase of input (onset I-O curve) changes in a rather subtractive way with increased levels of pre-adaptation [Benda2010]. This does not conflict the results shown here. In fact, it holds for both types of adaptation current (not shown).
% OMIT PERHAPS:
%However, the I-O curve can also be changed by other somatic components, such as the the leak current and sodium channel inactivation \citep{Fernandez2010}.
%
%%%Implications for gain control:
%%From Rothman2009:
%Gain modulation is a widespread neuronal phenomenon that modifies response amplitude without changing selectivity amplifying or scaling down the sensitivity of the neuron to changes in its input.
%Such gain modulation occurs in vivo, during contrast invariance of orientation tuning translation-invariant object recognition and coordinate transformations, attentional scaling, and auditory processing.
Modulation of response gain is an important phenomenon, particularly in sensory neurons, because neuronal sensitivity to changes in the input is amplified or downscaled without changing input selectivity. A spike-dependent adaptation current thus represents a cellular mechanism by which this is achieved. For example, neuronal response gain increases during selective attention \citep{McAdams1999}. It has been shown \emph{in-vivo} that the neuromodulator acetylcholine (ACh) contributes substantially to attentional upregulation of spike rates \citep{Herrero2008}. 
Cholinergic changes of neuronal excitability and response gain \citep{Soma2012} in turn are likely produced via downregulation of slow $\mathrm{K}^+$-currents \citep{Madison1987,McCormick1992,Sripati2006}. 
%Attentional modulation of neuronal responses by acetylcholine [Herrero2008] explained by cholinergic changes of response gain [Soma2012,2013] via downregulation of potassium currents [Sripati2006, Madison1987, McCormick1992]
Together with our results, these observations suggest that excitability and response gain of cortical neurons are controlled by neuromodulatory substances through (de)activation of subthreshold and spike-triggered $\mathrm{K}^+$-currents, respectively.

%%%Synaptic currents:
We have further shown that external inhibitory synaptic inputs change the I-O curve subtractively, which is consistent with the results of a previous numerical study using a conductance based neuron model without consideration of noise \citep{Capaday2002}.
Recurrent synaptic (feedback) inhibition, which is a function of the neuron's spike rate, on the other hand, reduces the response gain. This is in agreement with the results obtained by \citep{Sutherland2009} based on IF type neurons subject to noisy inputs. 
Recent \emph{in-vivo} recordings from mouse visual cortex have shown that distinct types of inhibitory neurons produce these differential effects (i.e., subtractive and divisive changes of I-O curves) 
at their target neurons \citep{Wilson2012}. % (target: pyramidal neurons)
%This finding could be explained by our ... 
Functional connectivity analysis suggests that the inhibitory neurons which changed the I-O curve of their target neurons subtractively were less likely connected recurrently to the recorded targets %(like external inhibition), 
than the inhibitory neurons which changed their targets' responses divisively \citep{Wilson2012}. %...produced a divisive change were more likely recurrently connected to the recorded one (like recurrent inhibition). This is consistent with our results / prediction ...
Applying our results based on the simple network model the observed differential effects caused by the two types of inhibitory cells can thus be explained by their patterns of connectivity with the target cells. \\ %or: Togehter with our ... this can explain the two distinct effects observed.

% OMIT PERHAPS:
%Moreover, gain modulation can also be mediated by synaptic plasticity, such as short-term depression, which has been shown to promote inhibition-induced divisive changes of I-O curves \citep{Rothman2009}. \\
%%Cardin2008? (cat V1 in-vivo)

\noindent \textbf{Effects on ISI variability} \\
We have shown that a spike-triggered adaptation current reduces high ISI variability at low spike rates (when input fluctuations are strong compared to the mean) and increases low ISI variability at high spike rates (caused by a large mean input). This result is in agreement with a previous numerical simulation study \citep{Liu2001} but seems to disagree with other theoretical work \citep{Wang1998,Prescott2008,Schwalger2010} at first sight. Wang (1998) and later Prescott \& Sejnowski (2008) showed that spike-driven adaptation reduces the ISI CV at low spike rates but they did not find an increase of ISI CV at higher spike rates in their simulation studies. The reason for this is that the ISI CVs of adapting and non-adapting neurons were compared at equal spike rates (i.e., at equal mean ISIs) but different input statistics. That is, the input to the adapting neurons was adjusted to compensate for the change of spike rate (or mean ISI) caused by the adaptation currents.
Increasing the mean input to the adapting neurons to achieve equal mean ISIs, however, decreases its ISI CV (cf. eq.~\eqref{cvISI_aPIF_results}). Here we compare the ISI statistics across different neurons for equal inputs. On the other hand, Schwalger et al. (2010) analyzed the ISI statistics of perfect IF model neurons with spike-triggered adaptation and found that this type of adaptation always leads to an increase of ISI CV in response to a noisy input current. Their approach is similar to the one presented here but differs in that the dynamics of the adaptation current was neglected in \citep{Schwalger2010}, see Fig.~\ref{fig0}B (bottom panel) for a visualization of that difference. Assuming a stationary adaptation current leads to a reduced effective mean input to the neuron, leaving the input variance unchanged, which always causes increased ISI variability, cf. eq.~\eqref{cvISI_aPIF_results}. 
%Our result above is further supported by in-vitro experimental data indicating that $\mathrm{Ca}^{2+}$-activated high-threshold slow $\mathrm{K}^+$-currents play an important role in controlling ISI variability \citep{Hallworth2003,Miles2005}.
Together with theoretical work showing that a spike-dependent adaptation current causes negative serial ISI correlations \citep{Prescott2008,Farkhooi2011} our results suggest that spike rate coding is improved by such a current for low-frequency inputs \citep{Prescott2008,Farkhooi2011}.
%Here, we have focused on the effects on different types of adaptation current and whether synaptic inhibition causes similar changes in ISIs. 

In contrast, an adaptation current which is predominantly driven by the subthreshold membrane voltage usually leads to an increase of ISI CV, as we have demonstrated. 
%This is achieved by a reduction of the effective mean input in a spike independent manner, while leaving the input variance unaffected, cf. eq.~\eqref{cvISI_aPIF_results}. 
This seems to be not consistent with a previous study \citep{Prescott2008} where subthreshold adaptation was found to produce a small decrease of ISI variability. The apparent discrepancy is caused by differences in the presentation of the data: Prescott \& Sejnowski (2008) compared the ISI CVs for equal spike rates as explained above. That is, the mean input was adjusted to obtain equal mean ISIs for adapting and non-adapting neurons but the input variance remained unchanged. However, increasing the mean input ($\mu$ in eq.~\eqref{cvISI_aPIF_results}) to the adapting neuron counteracts the effect of subthreshold adaptation on the effective mean input ($\mu_a$ in eq.~\eqref{cvISI_aPIF_results}). 
Consequently, one cannot observe an increased ISI CV in neurons with subthreshold adaptation currents when the mean input to these neurons is increased. 
Note that our results do not contradict those in \citep{Prescott2008}, but instead reveal that an increase of a subthreshold adaptation current always causes an increase of ISI CV for given input statistics and an increase of a spike-dependent adaptation current leads to an increase of ISI CV if the mean input is large. 
%This suggests that spike-dependent adaptation and synaptic excitation are indicative of a spike-rate code, while subthreshold adaptation and synaptic inhibition could promote coding schemes based on spike timing [Prescott2008]. 
%Lundstrom2006?

Finally, we have shown that an increase in synaptic inhibition increases the ISI variability, regardless of whether this inhibition originates from an external population of neurons or from recurrently coupled ones. An intuitive explanation for this effect is that increased inhibitory input reduces the mean input but increases the input variance, see eqs.~\eqref{mu_eq}--\eqref{sigma_eq}. 
\textcolor{black}{The reason why recurrent synaptic inhibition and spike-triggered adaptation change the ISI variability in opposite ways in a fluctuation-dominated input regime could be the different timescales. Synaptic inhibition usually acts on a much faster timescale than adaptation currents whose time constants range from about one hundred milliseconds to seconds. 
%The difference in the changes of ISI variability caused by recurrent synaptic inhibition and spike dependent adaptation in a fluctuation-dominant input regime could be explained by the difference in timescales. 
Thus, recurrent synaptic inhibition in contrast to spike-triggered adaptation cannot provide a memory trace of past spiking activity (over a duration of several ISIs) that could shape the ISI distribution.
Notably, our results on ISIs in a network setting} strictly apply to networks in asynchronous states. 
Recurrent synaptic inhibition, however, can also mediate oscillatory activity \citep{Brunel2000,Brunel2003,Isaacson2011,Augustin2013} where the variability of ISIs might be affected differently. % \citep{Brunel2000}.  

%% file: section_acknowledgements.tex
\section*{Acknowledgements}

This work was supported by DFG in the framework of collaborative research center SFB910. We thank Maziar Hashemi-Nezhad for helpful comments on the manuscript.

%% file: section_figures.tex
\begin{figure*}[ht!]
\centering
\includegraphics[width=0.75\textwidth]{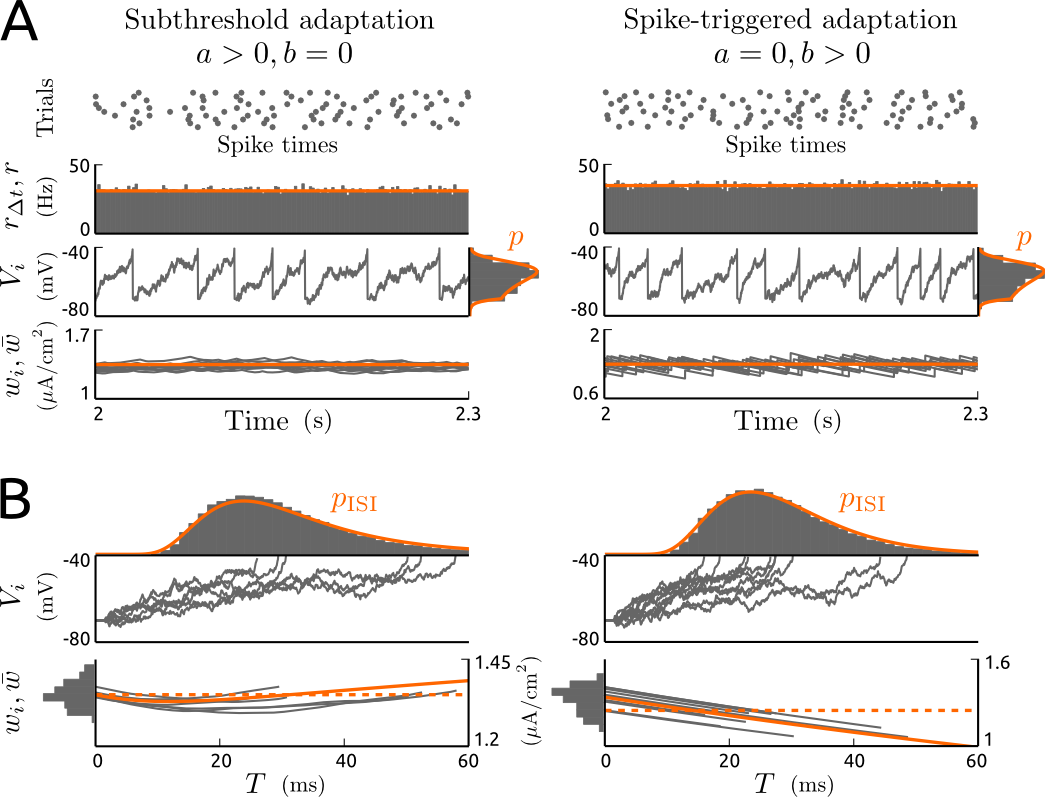} % removed \linewidth for doublespacing
\caption{\textbf{Steady-state spike rates and ISI distributions of single neurons.} 
A, from top to bottom: Spike times, instantaneous spike rate ($r_{\Delta t}$) histogram, membrane voltage ($V_i$), membrane voltage histogram and adaptation current ($w_i$) of an (adapted) aEIF neuron with $a=0.06$~mS/$\mathrm{cm}^2$, $b=0$ (left) and $a=0$, $b=0.18$~$\SI{}{\micro\ampere}/\mathrm{cm}^2$ (right) driven by a fluctuating input current with $\mu = 2.5$~mV/ms, $\sigma = 2$~mV/$\sqrt{\mathrm{ms}}$ for $N=5000$ trials. Spike times and adaptation current are shown for a subset of $10$ trials, the membrane voltage is shown for one trial. Results from numerical simulations are shown in grey. Results obtained using the Fokker-Planck equation are indicated by orange lines and include the instantaneous spike rate ($r$), the membrane potential distribution ($p$) and the mean adaptation current ($\bar{w}$). $r$, $p$ and $\bar{w}$ were calculated from the eqs.~\eqref{rate_eq}, \eqref{FP_eq} and \eqref{w_mean_eq}, respectively. These quantities have reached their steady state here. The time bin for $r_{\Delta t}$ was $\Delta t = 2$~ms, for the other parameter values see Materials and Methods.
B, top panel: ISI histogram corresponding to the $N$ trials in A and ISI distribution ($p_\mathrm{ISI}$, orange line) calculated via the first passage time problem (eq.~\eqref{pISI_eq}). B, center and bottom panels: Membrane voltage and adaptation current trajectories from one trial in A, but rearranged such that just after each spike the time is set to zero. Histograms for the adaptation current just after the spike times are included. The time-varying mean adaptation current from the first passage time problem (eq.~\eqref{w_mean_ISI_eq}) and the steady-state mean adaptation current from A (eq.~\eqref{w_mean_eq}) are indicated by solid and dashed orange lines, respectively.
All histograms (in A and B) represent the data from all $N$ trials.
}
\label{fig0}
\end{figure*}

\begin{figure*}[ht!]
\centering
\includegraphics[width=\textwidth]{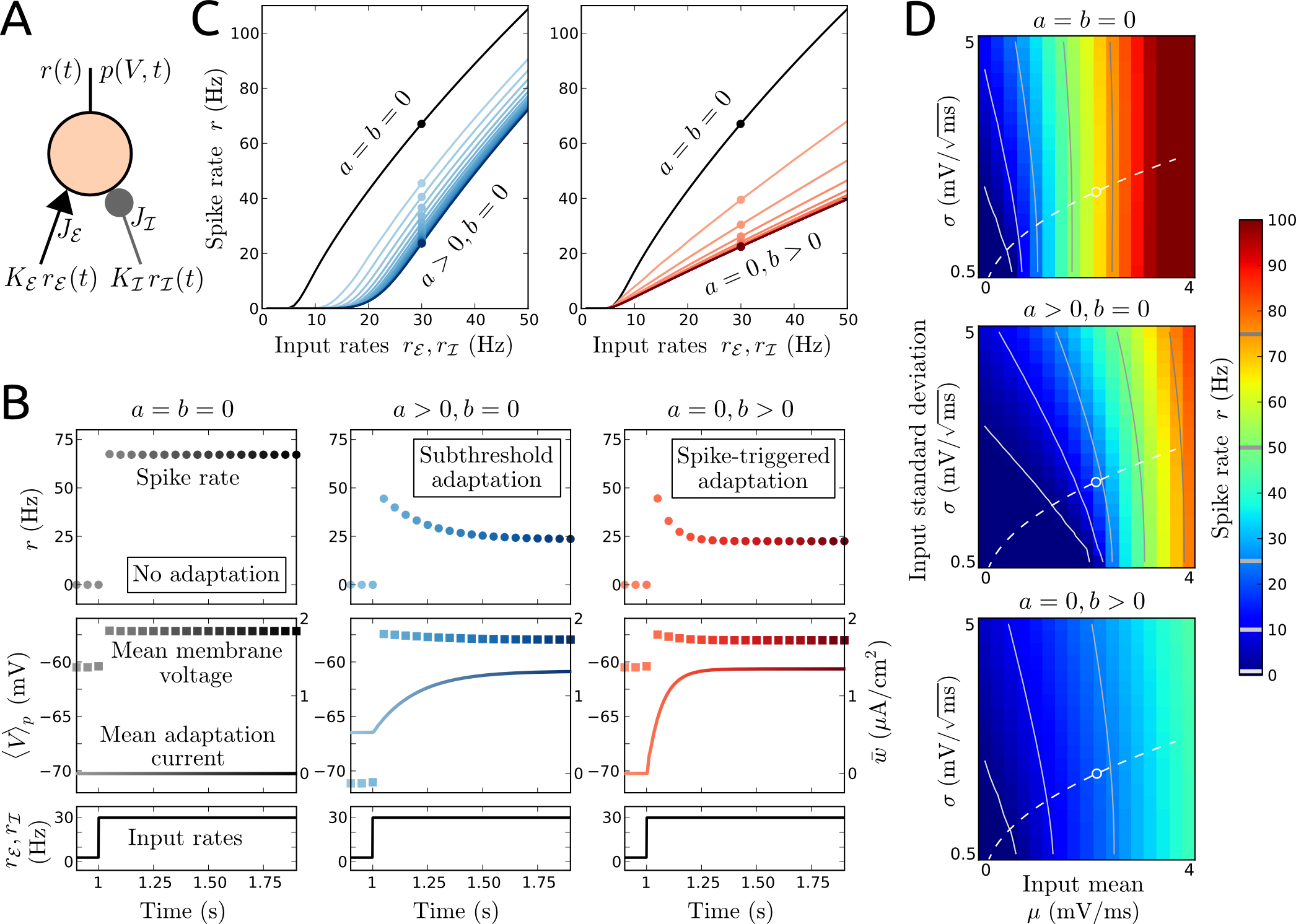} % removed \linewidth for doublespacing
\caption{\textbf{Spike rate adaptation, gain and threshold modulation in single neurons.} 
A: Cartoon of a single neuron visualizing the input parameters and output quantities.
B: Instantaneous spike rate $r$ (top panel), mean membrane voltage $\langle V \rangle_p$ (center panel, squares) and mean adaptation current $\bar{w}$ (center panel, solid lines) of an aEIF neuron without adaptation, $a=b=0$ (left), and with either a purely subthreshold adaptation current, $a=0.06$~mS/$\mathrm{cm}^2$, $b=0$ (center) or a spike-triggered adaptation current, $a=0$, $b=0.3$~$\SI{}{\micro\ampere}/\mathrm{cm}^2$ (right), in response to a sudden increase in synaptic drive (bottom panel). %That is, the presynaptic spike rates $r_\EE$ and $r_\II$ increase from $3$~Hz to $30$~Hz at $t=1$~s.
% 
%The neurons receive inputs from $K_\EE = 2000$ excitatory neurons which produce instantaneous PSPs of magnitude $J_\EE = 0.15$~mV and from $K_\II = 500$ inhibitory neurons producing instantaneous PSPs of magnitude $J_\II = -0.45$~mV. Each presynaptic excitatory and inhibitory neuron spikes with with Poisson rate $r_\EE$ and $r_\II$, respectively. Both, $r_\EE$ and $r_\II$ increase from $3$~Hz to $30$~Hz at $t=1$~s.  
%
C: I-O curves of the neurons in B, i.e., spike rate $r$ as a function of presynaptic spike rates $r_\EE$, $r_\II$. Here, $r_\EE = r_\II$, but excitation is stronger than inhibition, due to the coupling parameter values (see Materials and Methods). The I-O curves represent the spike rate response of the neurons to a sudden increase of $r_\EE$ and $r_\II$, measured in steps of $50$~ms after that increase (light to dark colors). Dots indicate the evolution of the spike rate corresponding to the input in B. 
D: Steady-state spike rate $r_\infty$ as a function of the mean $\mu$ and standard deviation $\sigma$ of the fluctuating input. Note that $\mu$ and $\sigma$ are determined by the number of presynaptic neurons, their (Poisson) spike rates and synaptic strengths, cf. eqs.~\eqref{mu_eq}--\eqref{sigma_eq}.
The dashed lines in D indicate the values of $\mu$ and $\sigma$ which correspond to the presynaptic spike rates in C, circles mark the values of the moments corresponding to the increased input in B. 
%TODO: Flip A,B, indicate $\mu$ and $\sigma$ in B, Dots in C,D, Visualize inputs in A
}
\label{fig1}
\end{figure*}

\begin{figure*}[ht!]
\centering
\includegraphics[width=\textwidth]{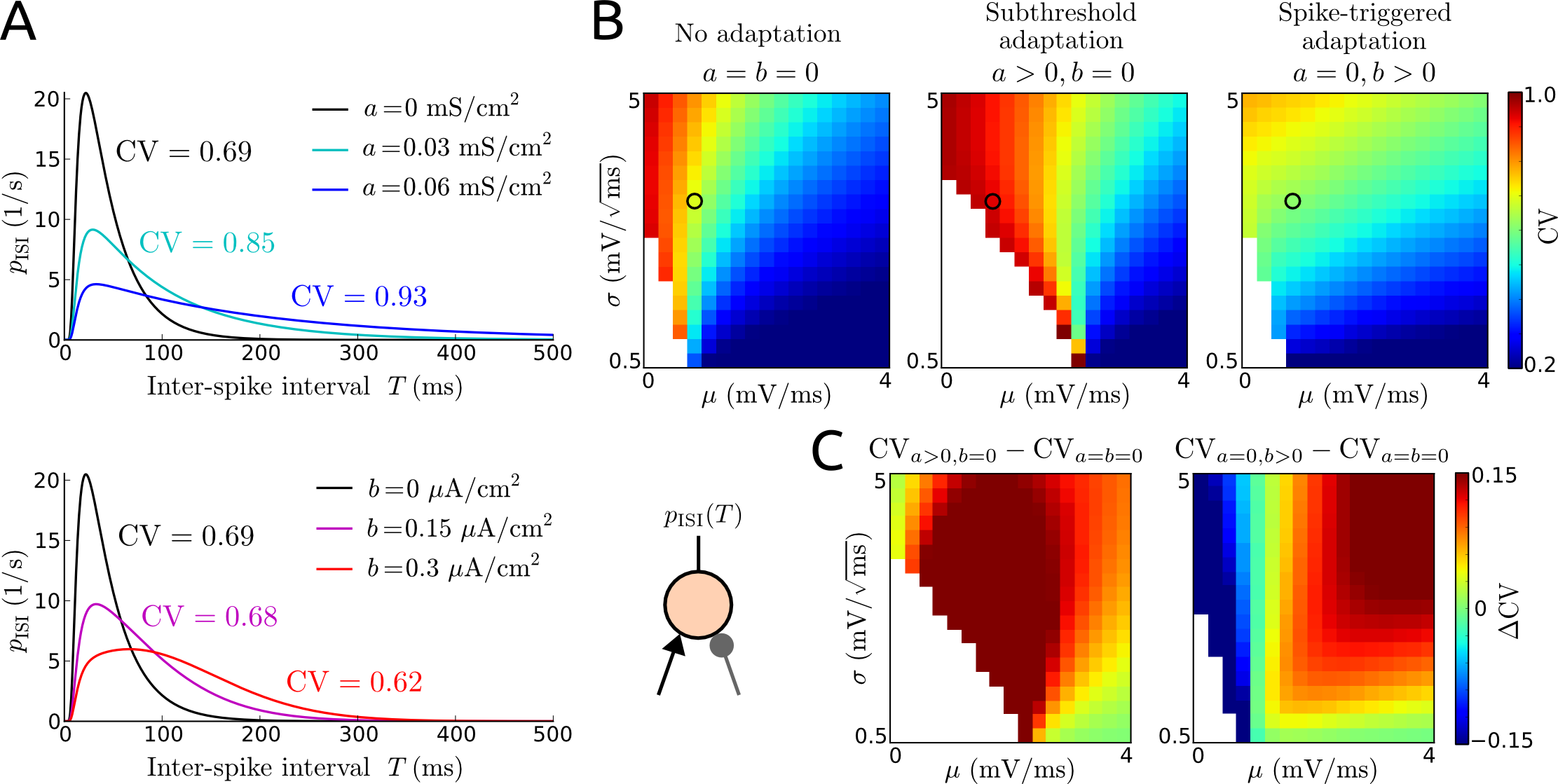}
\caption{\textbf{Changes of spiking variability in single neurons.} 
A: ISI distribution ($p_{\mathrm{ISI}}$) of a single aEIF neuron in response to a fluctuating input with mean $\mu=0.75$~mV/ms and standard deviation $\sigma=3.25$~mV/$\sqrt{\mathrm{ms}}$, for $a=0, 0.03, 0.06$~mS/$\mathrm{cm}^2$, $b=0$ (top) and $a=0$, $b=0, 0.15, 0.3$~$\SI{}{\micro\ampere}/\mathrm{cm}^2$ (bottom). 
B: ISI coefficient of variation (CV) as a function of $\mu$ and $\sigma$, for a neuron without adaptation, $a=b=0$ (left), and with either a subthreshold adaptation current, $a=0.06$~mS/$\mathrm{cm}^2$, $b=0$ (center) or a spike-triggered adaptation current, $a=0$, $b=0.3$~$\SI{}{\micro\ampere}/\mathrm{cm}^2$ (right). Circles indicate the values of $\mu$ and $\sigma$ used in A.
C: Change of ISI CV caused by a subthreshold (left) or spike-triggered (right) adaptation current as a function of $\mu$ and $\sigma$. The white regions in B and C indicate the parameter values for which the ISI CV was not computed, because $r_{\infty}<1$~Hz. 
%TODO: CVs in A, title in C, refine cartoon, Dots in B,C
}
\label{fig2}
\end{figure*}

\begin{figure*}[ht!]
\centering
\includegraphics[width=0.75\textwidth]{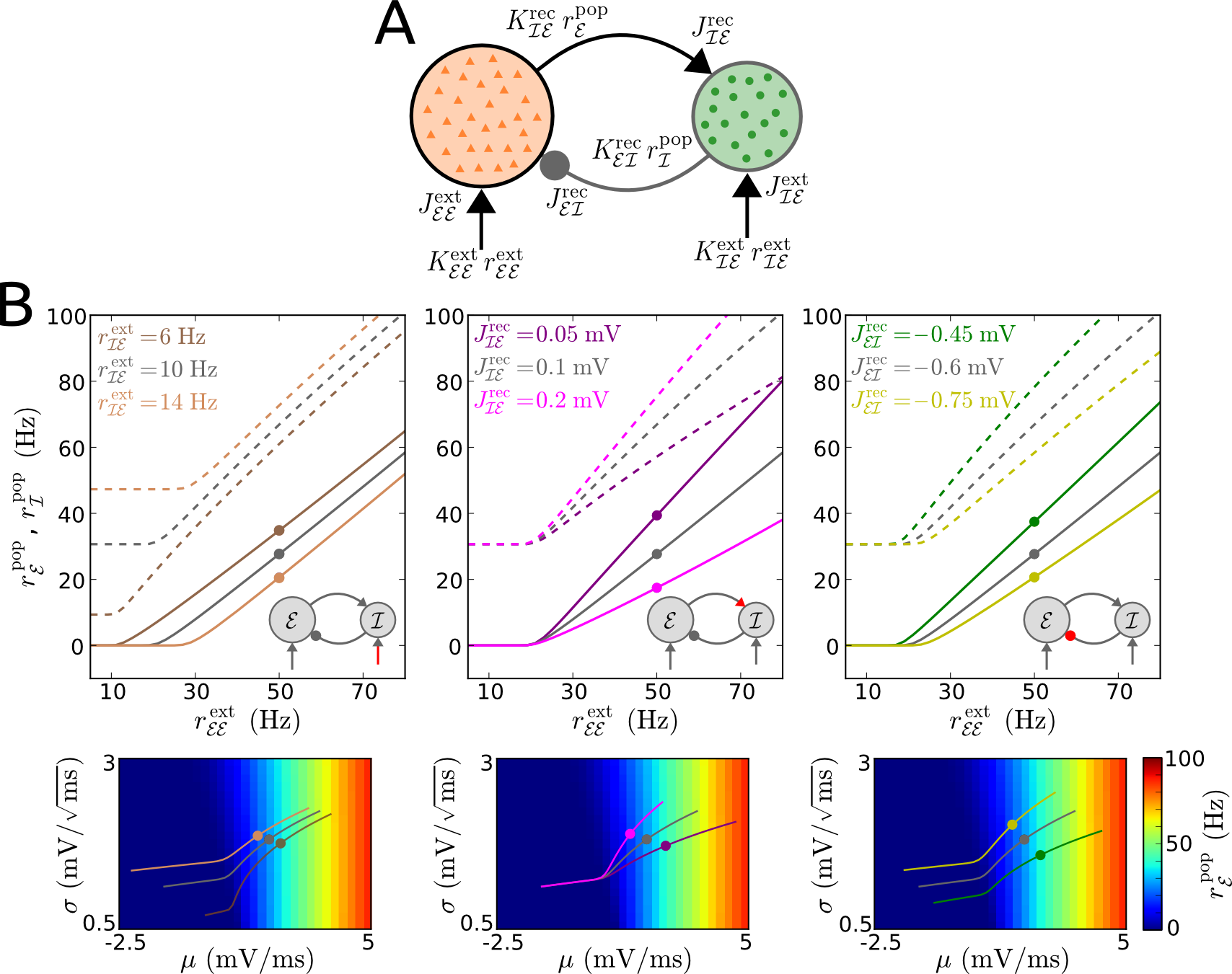}
\caption{\textbf{Gain and threshold modulation caused by network interaction.} 
A: Cartoon of the network visualizing the coupling parameters. 
%Each excitatory neuron receives inputs from $K_{\EE\EE}^{\mathrm{ext}} = 800$ external excitatory neurons with synaptic strength $J_{\EE\EE}^{\mathrm{ext}} = 0.15$~mV, spiking with rate $r_{\EE\EE}^{\mathrm{ext}}$, and from $K_{\EE\II}^{\mathrm{rec}} = 100$ inhibitory neurons of the network with synaptic strength $J_{\EE\II}^{\mathrm{rec}}$. 
%Each inhibitory neuron receives inputs from $K_{\II\EE}^{\mathrm{ext}} = 800$ external excitatory neurons with synaptic strength $J_{\II\EE}^{\mathrm{ext}} = 0.15$~mV, spiking with rate $r_{\II\EE}^{\mathrm{ext}}$, and from $K_{\II\EE}^{\mathrm{rec}} = 400$ excitatory neurons of the network with synaptic strength $J_{\II\EE}^{\mathrm{rec}}$. 
%MORE COMPACT, AFTER B;
B, top panel: Steady-state spike rate of excitatory aEIF neurons, $r_{\EE,\infty}^{\mathrm{pop}}$ (solid lines) and inhibitory aEIF neurons, $r_{\II,\infty}^{\mathrm{pop}}$ (dashed lines), as a function of $r_{\EE\EE}^{\mathrm{ext}}$, for 
$r_{\II\EE}^{\mathrm{ext}} = 6,10,14$~Hz (left),
$J_{\II\EE}^{\mathrm{rec}} = 0.05, 0.1, 0.2$~mV (center),
$J_{\EE\II}^{\mathrm{rec}} = -0.45, -0.6, -0.75$~mV (right).
Inset cartoons visualize the varied parameters as specified on the top left.
If not indicated otherwise, $J_{\EE\II}^{\mathrm{rec}} = -0.6$~mV, $r_{\II\EE}^{\mathrm{ext}} = 10$~Hz and $J_{\II\EE}^{\mathrm{rec}} = 0.1$~mV. For the other parameter values see Materials and Methods.
B, bottom panel: Steady-state spike rate $r_{\EE,\infty}^{\mathrm{pop}}$ as a function of the input parameters $\mu$ and $\sigma$ for the excitatory neurons. Solid lines and dots in the top panel correspond to those of equal color in the bottom panel.
%(No delay) 
%The adaptation parameter values were $a=0.015$~mS/$\mathrm{cm}^2$ , $b=0.1$~$\SI{}{\micro\ampere}/\mathrm{cm}^2$ for excitatory neurons and $a=b=0$ for inhibitory neurons, since adaptation has been shown to be weak for inhibitory interneurons [ref].
}
\label{fig3}
\end{figure*}

\begin{figure*}
\centering
\includegraphics[width=0.75\textwidth]{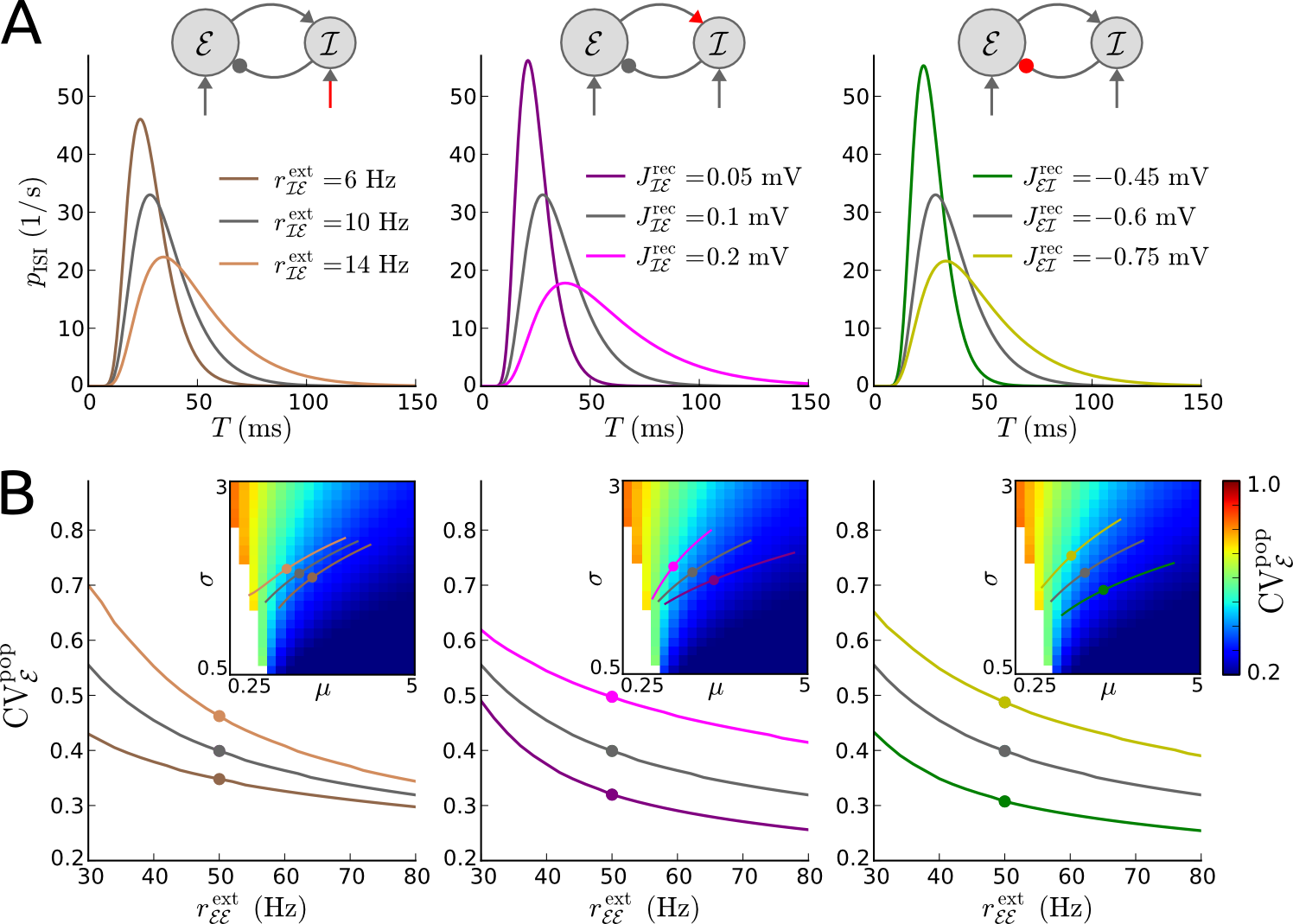}
\caption{\textbf{Changes of spiking variability caused by network interaction.} 
A: ISI distributions ($p_{\mathrm{ISI}}$) of excitatory aEIF neurons for $r_{\EE\EE}^{\mathrm{ext}} = 50$~Hz. $J_{\EE\II}^{\mathrm{rec}} = -0.6$~mV, $r_{\II\EE}^{\mathrm{ext}} = 10$~Hz and
$J_{\II\EE}^{\mathrm{rec}} = 0.1$~mV if not indicated otherwise. 
B: ISI CV for excitatory neurons ($\mathrm{CV}_{\EE}^{\mathrm{pop}}$) as a function of $r_{\EE\EE}^{\mathrm{ext}}$. Color code as in A. Dots indicate the input and ISI CV values for the ISI distributions in A. 
Insets: ISI CV as a function of the input parameters $\mu$ and $\sigma$ for the excitatory neurons. Lines and dots (insets) correspond to those of equal color in B.
% Other parameter values as in Fig.~\ref{fig3}. 
}
\label{fig4}
\end{figure*}

\begin{figure*}
\centering
\includegraphics[width=0.85\textwidth]{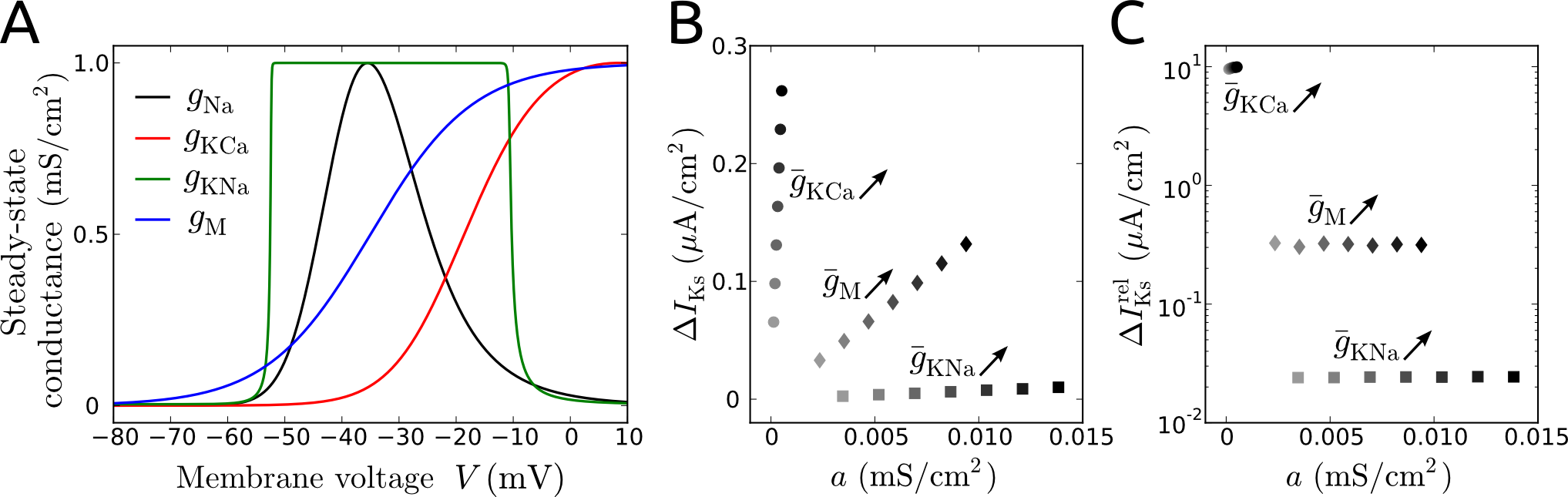}
\caption{\textbf{Subthreshold and spike-triggered components of $I_{\mathrm{KCa}}$, $I_{\mathrm{KNa}}$ and $I_{\mathrm{M}}$.} 
A: Conductances for the slow $\mathrm{K}^+$-currents $I_{\mathrm{Na}}$, $I_{\mathrm{KCa}}$, $I_{\mathrm{KNa}}$ and $I_{\mathrm{M}}$ in steady state as a function of the membrane voltage, normalized to a peak value of $1$~mS/$\mathrm{cm}^2$.
B and C: Subthreshold conductance $a$ and spike-triggered absolute increment $\Delta I_{\mathrm{Ks}}$ (B) and relative increment $\Delta I_{\mathrm{Ks}}^{\mathrm{rel}}$ (C) obtained from the fitting procedure (see Materials and Methods) for the conductance-based model neurons with 
$\bar{g}_{\mathrm{KCa}} \in [2,\, 8]$~mS/$\mathrm{cm}^2$ and $\bar{g}_{\mathrm{KNa}}=\bar{g}_{\mathrm{M}}=0$ (dots), 
$\bar{g}_{\mathrm{KNa}} \in [2,\, 8]$~mS/$\mathrm{cm}^2$ and $\bar{g}_{\mathrm{KCa}}=\bar{g}_{\mathrm{M}}=0$ (squares),
$\bar{g}_{\mathrm{M}} \in [0.1,\, 0.4]$~mS/$\mathrm{cm}^2$ and $\bar{g}_{\mathrm{KCa}}=\bar{g}_{\mathrm{KNa}}=0$ (diamonds). 
Darker symbols indicate larger conductance values.}
\label{fig5}
\end{figure*}

\begin{figure*}
\centering
\includegraphics[width=0.8\textwidth]{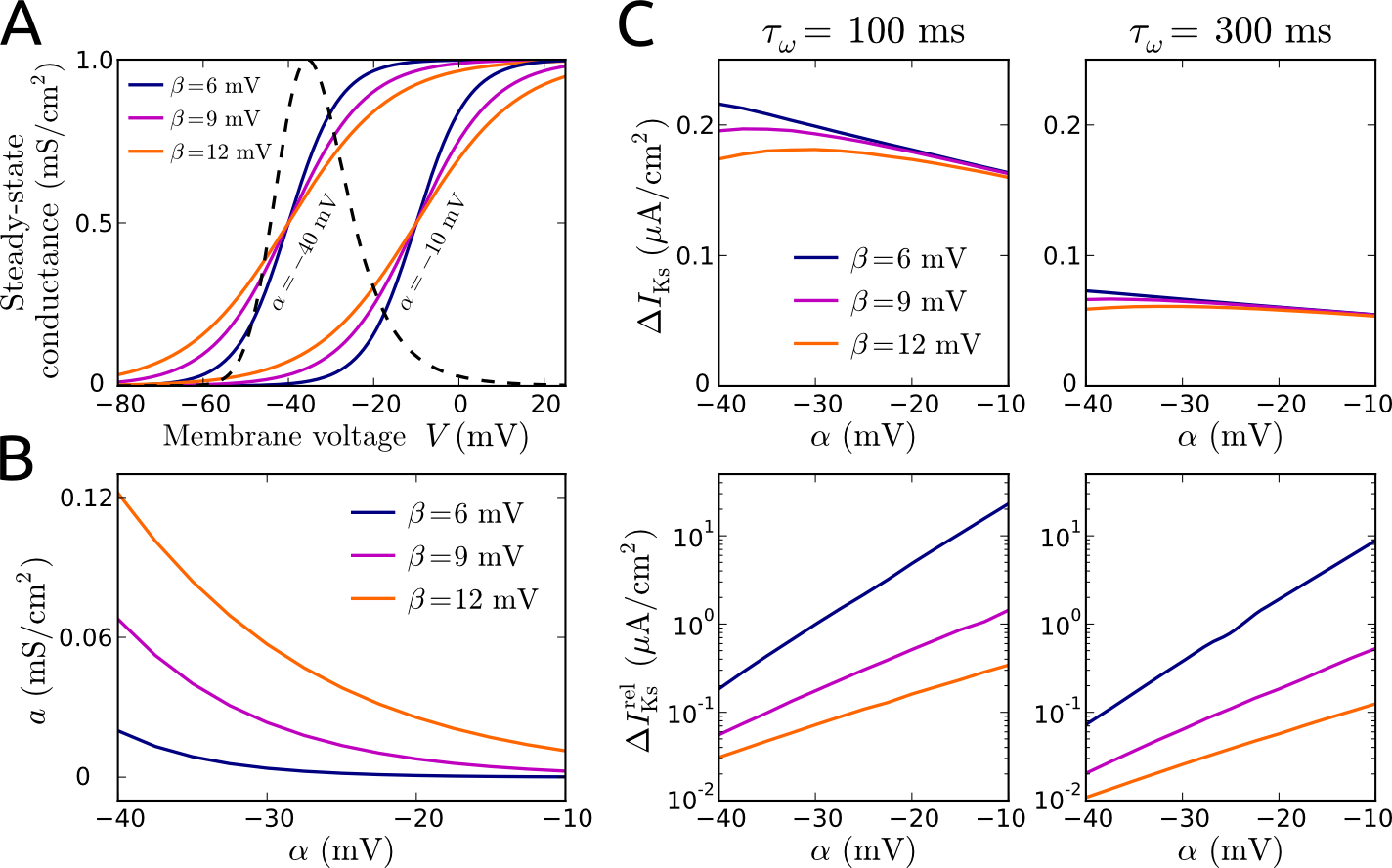}
\caption{\textbf{Subthreshold and spike-triggered components of a range of slow $\mathrm{K}^+$-currents}
A: Steady-state $\mathrm{K}^+$-conductance $g_{\mathrm{Ks},\infty}(V) = \bar{g}_{\mathrm{Ks}} \omega_{\infty}(V)$ as a function of the membrane voltage, for the generic Hodgkin-Huxley-type description of a slow $\mathrm{K}^+$-current (see Materials and Methods), with 
%$I_{\mathrm{Ks}}$, with steady-state conductance-voltage relationship $\omega_{\infty}(V)$ given by eq.~\eqref{HH_adapt_gating_eq} (see Materials and Methods) 
half-activation voltage $\alpha=-40$~mV (left curves), $\alpha=-10$~mV (right curves), inverse steepness $\beta=6,9,12$~mV and peak conductance $\bar{g}_{\mathrm{Ks}} = 1$~mS/$\mathrm{cm}^2$. The dashed curve indicates the $\mathrm{Na}^+$-conductance $g_{\mathrm{Na},\infty}(V)$ of the conductance-based model, normalized to a maximum value of $1$~mS/$\mathrm{cm}^2$.
B: Subthreshold conductance $a$ obtained from the fitting procedure for different values of the parameters $\alpha$ and $\beta$.
% $\alpha \in [-40, -10]$~mV and $\beta=6,9,12$~mV
C: Absolute and relative spike-triggered increments $\Delta I_{\mathrm{Ks}}$ (top panel) and $\Delta I_{\mathrm{Ks}}^{\mathrm{rel}}$ (bottom panel), respectively, as a function of $\alpha$, for 
% $\beta=6,9,12$~mV, 
$\tau_{\omega}=100$~ms (left) and $\tau_{\omega}=300$~ms (right).  
}
\label{fig6}
\end{figure*}

\begin{figure*}
\centering
\includegraphics[width=0.6\textwidth]{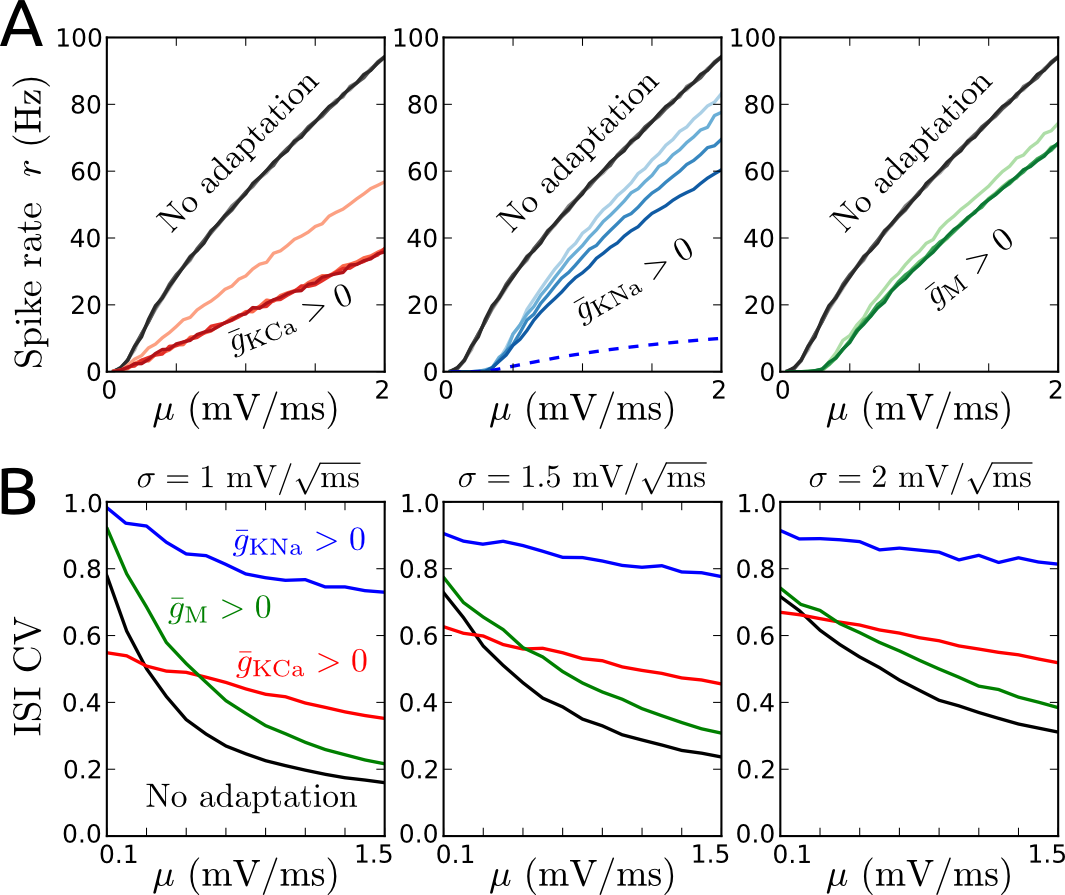}
\caption{\textbf{Effects of $I_{\mathrm{KCa}}$, $I_{\mathrm{KNa}}$ and $I_{\mathrm{M}}$ on I-O curve and ISI variability.} 
A: Spike rate of a conductance-based model neuron without slow $\mathrm{K}^+$-currents, $\bar{g}_{\mathrm{KCa}}=\bar{g}_{\mathrm{KNa}}=\bar{g}_{\mathrm{M}} = 0$ (black), and with either type of slow $\mathrm{K}^+$-current included, $\bar{g}_{\mathrm{KCa}}= 8$~mS/$\mathrm{cm}^2$ (red), $\bar{g}_{\mathrm{KNa}}= 8$~mS/$\mathrm{cm}^2$ (blue), $\bar{g}_{\mathrm{M}}= 0.4$~mS/$\mathrm{cm}^2$ (green), in response to a sudden increase of mean input $\mu$, measured in four subsequent time intervals of $250$~ms after that increase (light to dark colors). The baseline mean input was $\mu = 0.05$~mV/ms and the input standard deviation was $\sigma = 0.5$~mV/$\sqrt{\mathrm{ms}}$. Average values over $50$ independent trials are shown. % actually 20, maybe REPEAT with 50 for smoother curves 
The adapting I-O curve of the neuron with increased $I_{\mathrm{KNa}}$ ($\bar{g}_{\mathrm{KNa}}= 8$~mS/$\mathrm{cm}^2$) converges very slowly to the steady-state curve (dashed blue) measured $20$ s after the increase in $\mu$.
%or: The spike rate of the neuron with increased $I_{\mathrm{KNa}}$ 
%($\bar{g}_{\mathrm{KNa}}= 8$~mS/$\mathrm{cm}^2$) adapts very slowly; its steady-state I-O curve, measured $20$~s after the increase in $\mu$, is indicated by the dashed blue curve. 
B: ISI CV of the neurons in A as a function of mean input $\mu$ for low (left), medium (center), and high (right) noise intensity ($\sigma = 1, 1.5, 2$~mV/$\sqrt{\mathrm{ms}}$), respectively. The ISIs were collected over an interval of $10$~s after the steady-state spike rates were reached, in $50$ independent trials.
% mabye repeat using single neurons but longer interval, e.g. 50s
}
\label{fig7}
\end{figure*}